\documentclass[aps,prb,preprint,showpacs,superscriptaddress,amssymb,amsmath]{revtex4-1} 
\usepackage{latexsym}
\usepackage[pdftex]{graphicx}
\usepackage{bm}
\usepackage{epsfig}
\usepackage{dcolumn}

\begin{document}

\title{X-ray cross-correlation analysis applied to disordered
two-dimensional systems}

\author{R.P. Kurta}
\affiliation{Deutsches Elektronen-Synchrotron DESY, Notkestra\ss e 85, D-22607 Hamburg, Germany}
\author{M. Altarelli}
\affiliation{European X-ray Free-Electron Laser Facility, Notkestra\ss e 85, D-22607, Hamburg, Germany}
\author{E. Weckert}
\affiliation{Deutsches Elektronen-Synchrotron DESY, Notkestra\ss e 85, D-22607 Hamburg, Germany}
\author{I.A. Vartanyants}
\email[Reference author: ]{ivan.vartaniants@desy.de}
\affiliation{Deutsches Elektronen-Synchrotron DESY, Notkestra\ss e 85, D-22607 Hamburg, Germany}
\affiliation{National Research Nuclear University, ``MEPhI'', 115409 Moscow, Russia}
\date{\today}

\begin{abstract}

Angular x-ray cross-correlation analysis (XCCA) is an approach to study the structure of disordered systems
using the results of coherent x-ray scattering experiments.
Here, we present the results of simulations that validate our theoretical findings for XCCA
obtained in a previous paper [M. Altarelli \textit{et al.}, Phys. Rev. B 82, 104207 (2010)].
We consider as a model two-dimensional (2D) disordered systems composed of non-interacting colloidal clusters with fivefold symmetry and with orientational and positional disorder. We simulate a coherent x-ray scattering in the far field from such disordered systems and perform the angular cross-correlation analysis of calculated diffraction data. The results of our simulations show the relation between the Fourier series representation of the cross-correlation functions (CCFs) and different types of correlations in disordered systems. The dependence of structural information extracted by XCCA on the density of disordered systems and the degree of orientational disorder of clusters is investigated. The statistical nature of the fluctuations of the CCFs in the model `single-shot' experiments is demonstrated and the potential of extracting structural information from the analysis of CCFs averaged over a set of diffraction patterns is discussed. We also demonstrate the effect of partial coherence of x-rays on the results of XCCA.

\end{abstract}


\maketitle


\section{Introduction}

The study of angular correlations in diffraction patterns has a long history with two main directions. On one hand, it has been used in small angle x-ray scattering experiments as a tool in the attempt to solve structures of molecules in solutions or, more generally, in non-periodic systems. It goes back to the work of Kam\cite{Kam, Kam1}, over 30 years ago; in much more recent times, the perspectives opened by ultra-short and ultra-bright pulses from free-electron lasers\cite{Ackermann, Emma, Tanaka5, Altarelli2}, which provide an opportunity to acquire diffraction data in a time short compared to translational and rotational relaxation times, have rekindled interest in this approach \cite{Saldin, Saldin1}.
In these recent applications, the angular correlation is defined as an average over many diffraction patterns.

An alternative purpose, which may seem more modest at first, but could be of paramount importance in the physics of disordered or partially ordered systems is the unveiling of hidden symmetries in a disordered collection of elements. This leads to the problem of understanding systems with the so-called medium range order\cite{Elliott, Sheng, Treacy} as well as complicated dynamics and correlation between dynamical heterogeneity and medium-range order in a large class of glass-forming liquids\cite{Tanaka3, Tanaka2, Tanaka1, Rice1, Rice2}. In such systems the relevant question will be, for example, can one recognize and identify an $n$-fold symmetry axis of an individual molecular species from the diffraction patterns of a liquid composed of such molecules? Can bond angles be detected from the study of angular correlations of the diffracted intensity of an amorphous system? There are partly affirmative answers in the study of partially ordered quasi two-dimensional systems like liquid crystals\cite{Pindak} in which hexatic bond order can be detected by the study of the angular dependence of  the diffracted intensity. More recently Wochner \emph{et al.}\cite{Wochner1, Wochner2} reported angular correlations with pronounced periodic character in a colloidal suspension of polymethylmethacrylate (PMMA) spheres, expected to form icosahedral clusters near the glass formation concentration.

In our first paper\cite{Altarelli}, angular correlations in the diffraction patterns from a disordered collection of symmetric species were discussed theoretically in a general frame. The information content of angular correlation data in different experimental conditions (dilute versus dense systems, different statistics of orientational disorder, and so on) was derived. It transpires that generally the richest information is obtainable from dilute systems in which not too many molecules or clusters are found instantly in the illuminated volume. At the same time for such systems, static fluctuations of the scattered intensity are relatively high, and do not give confirmative information on the local structure of disordered systems. It also appears that partial orientation of the species, with lining up of the symmetry axes in one direction, enhances the effects considerably, in agreement with the results of other groups. Nonetheless, some aspects of the data by Wochner \emph{et al.}\cite{Wochner1} were difficult to reconcile with our results, for example the pronounced fivefold symmetry of some of the observed angular correlation patterns.

In this paper we apply the x-ray cross-correlation analysis (XCCA) to the study of the local structure (LS) of disordered systems. In general, the role of LSs can be played by clusters in the gas phase or a colloidal system, protein molecules, viruses, or complex biological systems in solution. Here, we consider the particular case of model 2D systems composed of non-interacting clusters with identical shape. The main goal of this paper is to identify experimental conditions when symmetry and/or structure of clusters composing a disordered system can be correctly determined from the XCCA. By varying the density of clusters and characteristics of their orientational disorder, we analyze how these changes influence the angular cross-correlation functions (CCFs) and their Fourier spectra. We demonstrate a statistical behavior of CCFs calculated for different single realizations of a system and compare them with CCFs averaged over many realizations of the system (ensemble of diffraction patterns). Such analysis can be related to so-called 'single-shot' experiments, when the measurement time is much shorter than any typical structural relaxation time of the system under investigation. Such type of experiments can be performed, for example, on fourth generation x-ray sources, such as free-electron lasers\cite{Ackermann, Emma, Tanaka5, Altarelli2} (FELs), that produce ultrashort pulses on the femtosecond scale. We also investigate the influence of partial coherence of the incident x-ray beams on the outcome of the XCCA.

This paper is organized as follows. In the next section a brief theoretical description of the XCCA based on our previous work\cite{Altarelli} is given.
In the third section a model of a coherent scattering experiment on a disordered system used in simulations is introduced.
In the fourth section we study the CCF as a function of orientational disorder in the systems with a large number of clusters. Statistical fluctuations of the CCF in a model `single-shot' experiment for completely disordered systems with a small number of clusters are analyzed in the fifth section. Results of calculations of the averaged CCFs for systems with different number of particles and degree of orientational disorder are presented in the sixth section. In the seventh section we demonstrate the role of coherence properties of the incoming x-ray wavefields for the analysis of the local structure of disordered systems.
The paper is completed by the conclusions and outlook section. Detailed derivations are summarized in the Appendix section.

\section{Theoretical background}
The intensity CCF $C_{q}(\Delta)$ as a function of the angular coordinate $0\leq\Delta\leq 2\pi$ is defined as follows
\footnote{In this paper we consider, for simplicity, the reduced form of the CCF $C_{q_{1},q_{2}}(\Delta)$ introduced in Ref.~\cite{Altarelli} with $q_{1}=q_{2}=q$. Note, that in some cases, for example in Ref.\cite{Saldin1}, a not normalized form of the CCF $C_{q_{1},q_{2}}(\Delta)$ is used. In this case the influence of the form-factors of the individual scatterers on the form of CCFs is not excluded by normalization.}
\begin{equation}
C_{q}(\Delta) =
\frac{\left\langle I(q,\varphi)I(q,\varphi+\Delta)\right\rangle _{\varphi} - \left\langle I(q,\varphi)\right\rangle_{\varphi}^{2}}
{\left\langle I(q,\varphi)\right\rangle_{\varphi}^{2}},
\label{Eq:Cq1q2_0}
\end{equation}
where $I(q,\varphi)$ is the intensity scattered at the momentum transfer vector $\mathbf{q}$,
$\varphi$ is an angular coordinate around a diffraction ring of radius $q$, and
$
\langle f(\varphi) \rangle_{\varphi}=1/2\pi\int_{0}^{2\pi}f(\varphi)d\varphi
$
denotes the angular average around the ring.

It is convenient to analyze the CCF $C_{q}(\Delta)$ using a Fourier series decomposition\cite{Altarelli},
\begin{subequations}
\begin{eqnarray}
&&C_{q}(\Delta)=2\sum\limits_{n=1}^{\infty} C_{q}^{n}\cos(n\Delta),\label{Eq:Cq1q2_1}\\
&&C_{q}^{n}=\frac{1}{\pi}\int_{0}^{\pi}C_{q}(\Delta)\cos(n\Delta),\label{Eq:Cq1q2n_1}
\end{eqnarray}
\end{subequations}
where $C_{q}^{n}$ are the Fourier components.

The Fourier components of the CCF $C_{q}(\Delta)$ for $n\neq 0$ are determined by the angular
distribution of intensity $I(q,\varphi)$ according to
the following relation\cite{Altarelli}
\begin{equation}
C_{q}^{n}=\left\vert I^{n}(q)/I^{0}(q) \right\vert^{2},\label{Eq:Cq1q2_2}
\end{equation}
where $I^{n}(q)=1/({2\pi})\int_{0}^{2\pi}I(q,\varphi)e^{-in\varphi}d\varphi$ are the components of the Fourier expansion of the intensity on the ring of radius $q$. According to its definition the zero Fourier component of the intensity distribution $I^{0} (q)$ coincides with the small angle x-ray scattering (SAXS) form-factor $S(q)$. We also want to note here that normalization in Eq. (\ref{Eq:Cq1q2_2}) has to be done with caution, the $q-$values where $I^{0} (q) = 0$ should be excluded from the consideration.

In the following we will consider a specific case of two-dimensional (2D) disordered systems that on one hand simplifies XCCA analysis but at the same time reveals important general features common for such systems.
It was shown\cite{Altarelli}, that in this case only even ($n=2l, l=1,2,3,\dots$) Fourier components of the intensity $I^{n}(q)$  are giving non-zero contributions and they can be presented as a sum of two terms, attributed to the different types of structural correlations in the system
\begin{eqnarray}
{I}^{n}({q}) =(i)^{n} \left[\sum\limits _{k=1}^{N} L_{k}^{n}({ q})+ 2\cdot
\sum\limits_{k_{2}> k_{1}}L_{k_{1},k_{2}}^{n}({ q})\right],\quad n=2l.
\label{Inq}
\end{eqnarray}
Here the first sum over $k_{1}=k_{2}=k$ contains the contributions $L_{k}^{n}({ q})$ from $N$ individual clusters, and the second sum over $k_{1}\neq k_{2}$ contains the cross-terms $ L_{k_{1},k_{2}}^{n}({ q})$ corresponding to correlation of different clusters.
In 2D systems these two terms $L_{k}^{n}({ q})$ and $L_{k_{1},k_{2}}^{n}({ q})$
can be presented in the following way\cite{Altarelli}
\begin{subequations}
\begin{eqnarray}
&&L_{k}^{n}({q}) = \int\int d\mathbf{ r}_{1} d\mathbf{ r}_{2} \widetilde{\rho}_{k}(\mathbf{ r}_{1} )\widetilde{\rho}_{k}(\mathbf{ r}_{2} )J_{n}({q}|\mathbf{ r}_{21} |)e^{-in\phi_{\mathbf{ r}_{21} }},\label{Lnk}\\
&&{ L}_{k_{1},k_{2}}^{n}({ q}) =\int\int d\mathbf{ r}_{1} d\mathbf{ r}_{2}\widetilde{\rho}_{k_{1}}(\mathbf{ r}_{1})\widetilde{\rho}_{k_{2}}(\mathbf{ r}_{2})J_{n}(q|\mathbf{ R}_{k_{2},k_{1}}^{21}|)e^{-in\phi_{\mathbf{ R}_{k_{2},k_{1}}^{21}}}.
\label{Ln_k1k2}
\end{eqnarray}
\end{subequations}
Here $\mathbf{ R}_{k_{2},k_{1}}^{21}=\mathbf{ R}_{k_{2},k_{1}}+\mathbf{ r}_{21}$, where $\mathbf{ R}_{k_{2},k_{1}}$ is the 2D vector connecting centers of different clusters,  $\mathbf{ r}_{1}$ and $\mathbf{ r}_{2}$ are the 2D vectors defined in the local coordinate system (with its origin located at the center of $k$-th cluster). In Eqs.~(\ref{Lnk}, \ref{Ln_k1k2}) $\phi_{\mathbf{ r}_{21} }$ and $\phi_{\mathbf{ R}_{k_{2},k_{1}}^{21}}$ are the angles of the vectors $\mathbf{ r}_{21}=\mathbf{ r}_{2}-\mathbf{ r}_{1}$ and $\mathbf{ R}_{k_{2},k_{1}}^{21}$ in the external coordinate system,
 $\widetilde{\rho_k}(\mathbf{ r})$ is a projected electron density of a cluster, $J_{n}(qr)$ is the Bessel function of the first kind of integer order $n$, and the integration is performed over the area of a cluster.
 
If all clusters in the system have the same internal structure but are randomly oriented in 2D space,
the phase $\phi_{\mathbf{ r}_{21} }$ in the exponent of Eq.~(\ref{Lnk}) can be defined as $\phi_{\mathbf{ r}_{21} }=\phi_{k}+\phi_{\mathbf{ r}_{21}}^{0}$, where $\phi_{k}$ is the rotation angle of the $k$-th cluster with respect to the fixed (reference) angular orientation $\phi_{\mathbf{ r}_{21}}^{0}$ of the cluster\cite{Altarelli}. In this case,
the term $L_{k}^{n}({ q})$ can be expressed as follows
\begin{equation}
L_{k}^{n}(q)=e^{-in\phi_{k}}{  L}^{n}(q),
\label{LSepar}
\end{equation}
and the integral $L^{n}(q)$ is the same for all clusters:
\begin{equation}
L^{n}(q)=\int\int d\mathbf{ r}_{1} d\mathbf{ r}_{2} \widetilde{\rho}(\mathbf{ r}_{1} )\widetilde{\rho}(\mathbf{ r}_{2} )J_{n}(q
|\mathbf{ r}_{21} |)e^{-in\phi_{\mathbf{ r}_{21}}^{0}}.
\label{LocStr1}
\end{equation}

It is well seen from the structure of the integral in Eq.~(\ref{LocStr1}) that there is a direct correspondence between the electron density $\widetilde{\rho}(\mathbf{r})$ of LSs and the values of $L^{n}(q)$. If $L^{n}(q)$ is known, certain information about the electron density $\widetilde{\rho}(\mathbf{r})$, for example symmetry of clusters, can be obtained. Importantly, the first term in decomposition (\ref{Inq}) does not depend on inter-particle distance and will have the {\it same} contribution for dilute and dense systems. Contrary to that, according to (\ref{Ln_k1k2}) the term
${ L}_{k_{1},k_{2}}^{n}({ q})$ contains the inter-particle distance $\mathbf{ R}_{k_{2},k_{1}}$ as an important parameter. Clearly, the contribution of this term will be different for dilute and dense systems.

In the case of a \textit{dilute} disordered system the contribution of the second sum in Eq.~(\ref{Inq}) can be neglected\cite{Altarelli}, and substituting Eqs.~(\ref{Inq}) and (\ref{LSepar}) in Eq.~(\ref{Eq:Cq1q2_2}) we obtain for the Fourier components $C_{q}^{n}$ of the CCF
\begin{equation}
C_{q}^{n}=\mathcal{L}^{n}(q)\cdot A^2_{n},
\label{CnqDilute}
\end{equation}
where
\begin{equation}
\mathcal{L}^{n}(q)= \left\vert L^{n}(q)/L^{0}(q) \right\vert^{2},
\label{Lnq}
\end{equation}
is the structural term, which is directly related to the structure of a LS, and
$A_{n}$ is the statistical term, which depends on the orientational distribution of LSs in the system. It is defined as an amplitude of the random phasor sum\cite{Goodman1,Goodman2},
\begin{equation}
{\mathbf A}_{n}=A_{n} e^{i\theta_{n}}=\frac{1}{N}\sum\limits_{k=1}^{N} e^{in\phi_{k}},
\label{2.24}
\end{equation}
where $\theta_{n}$ is the phase of this sum.
It is clear from Eq.~(\ref{CnqDilute}) that the value of the Fourier component $C_{q}^{n}$  in dilute systems is determined
by the product of the structural and statistical terms.

We analyzed the statistical behavior of $A^2_{n}$ for different angular distributions of cluster orientations
in our previous paper\cite{Altarelli}. Our analysis has shown that
in the case of a completely \textit{oriented} dilute system (all $\phi_{k}=\phi_{0}$), the statistical term is equal to unity,
and the Fourier components reach their maximum values, determined by the structural term $C_{q}^{n}=\mathcal{L}^{n}(q)$.
If the distribution of orientations of LSs is \textit{Gaussian}, with the standard deviation $\sigma_\phi$ and a zero mean $\langle\phi\rangle = 0$, the average value of the statistical term is
\begin{equation}
\langle A_{n}^2\rangle = \exp{\left(- n^{2}\sigma^{2}_\phi\right)} \left(1-1/N\right) + 1/N.
\label{GausDistr}
\end{equation}
It follows from Eq.~(\ref{GausDistr}), that in the limit $N\to\infty$, the Fourier components 
of the averaged angular CCF are defined by [see for details section VI]
\begin{equation}
\langle C_{q}^{n} \rangle = \mathcal{L}^{n}(q) \exp{\left(- n^{2}\sigma^{2}_{\phi}\right)}.
\label{2.26}
\end{equation}
According to these results for a dilute system with Gaussian distribution of orientations, the contribution of
higher order Fourier components (corresponding to large $n$-values) will be much lower compared
to the contribution of the lower order Fourier components.

In the case of a \textit{uniform} distribution of a large number $N$ of orientations, the statistical term fluctuates around its mean value $\langle A_{n}^{2} \rangle$ with the standard deviation $\sigma_{A_{n}^{2}}$:
\begin{equation}
\langle A_{n}^{2} \rangle = 1/N,\quad\sigma_{A_{n}^{2}}=1/N.
\label{UnifDistr}
\end{equation}
It means that in this case fluctuations are of the same order as the mean value and their ratio does not depend on the number of clusters in the system. According to Eqs.~(\ref{UnifDistr}) and (\ref{CnqDilute}), in the limit of a large number of
orientations ($N \to \infty $) the Fourier components of the averaged CCF have vanishing values, $\langle C_{q}^{n}\rangle \to 0$.

\bigskip

In the case of a \textit{dense} disordered system, when the average distance between LSs is of the order of the size of a single cluster, the second sum in Eq.~(\ref{Inq}) can not be neglected. Importantly, for such a dense system it can significantly affect the Fourier spectrum of the angular CCF. Taking into account both terms of Eq.~(\ref{Inq}), the Fourier components of the angular CCF (\ref{Eq:Cq1q2_2}) can be written as the following sum
\footnote{Note, that in Eq.~(\ref{Cq1q2n_2}) we combined the complex-valued terms 2 and 3 from Ref.~\cite{Altarelli} into one real-valued term.}
\begin{equation}
C_{q}^{n}=\left\vert \frac{I^{n}(q)}{I^{0}(q)}\right\vert^{2}=\frac{S^{n}_{1}+(S^{n}_{2}+S^{n}_{3})+S^{n}_{4}}{\left\vert I^{0}(q)\right\vert^{2}}
,\label{Cq1q2n_2}
\end{equation}
where
\begin{subequations}
\begin{eqnarray}
S^{n}_{1}&&=\left\vert \sum\limits_{k} L_{k}^{n}(q)\right\vert^2=N^2\left\vert L^{n}(q)\right\vert^2 A^{2}_{n},\label{Cq1q2n_T1}\\
S^{n}_{2}+S^{n}_{3}&&=4\cdot\text{Re}\left[ \sum\limits_{k} L_{k}^{n\ast}(q)\cdot\sum\limits_{k_{2}>k_{1}} L_{k_{1},k_{2}}^{n}(q)\right],\label{Cq1q2n_T23}\\
S^{n}_{4}&&=4\left\vert \sum\limits_{k_{2}>k_{1}} L_{k_{1},k_{2}}^{n}(q) \right\vert^2,\label{Cq1q2n_T4}
\end{eqnarray}
\end{subequations}
and
\begin{equation}
\left\vert I^{0}(q)\right\vert^{2}
=\left\vert NL^{0}(q)+2\sum\limits_{k_{2}>k_{1}} L_{k_{1},k_{2}}^{0}(q)  \right\vert^2.\label{Cq1q2n_I0}
\end{equation}
The summation in Eqs.~(\ref{Cq1q2n_T1}-\ref{Cq1q2n_T4}) is performed over all LSs.
In Eq.~(\ref{Cq1q2n_2}) the term $S^{n}_{1}$ is defined by the local structure, and terms $S^{n}_{2}, S^{n}_{3}$ and 
$S^{n}_{4}$ contain contributions from the inter-particle correlations due to the second term in Eq.~(\ref{Inq}).
 Below we will explicitly investigate contribution of these terms to the total value of the Fourier components $C_{q}^{n}$ of the CCF.


\section{Model of a disordered system and scattering geometry}


(a) {\it Model}

\bigskip

\begin{figure*}[!htbp]
\centering
\includegraphics[width=1.0\textwidth]{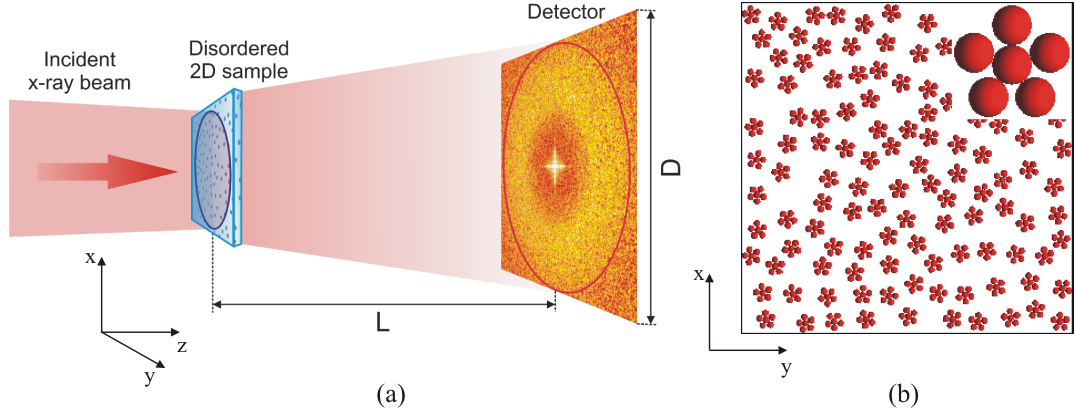}
\caption{\label{Fig:ExpGeometr}(Color online) (a) Geometry of the diffraction experiment. A coherent x-ray beam illuminates a 2D disordered sample and produces a diffraction pattern on a detector. The direction of the incident beam is defined along the z axis of the coordinate system. (b) A disordered 2D sample composed of centered pentagonal clusters (an enlarged view of the cluster is shown in the inset). The case when all clusters have random position and orientation in the 2D plane is shown.}
\end{figure*}

The model and the geometry of the coherent x-ray scattering experiment used in our simulations
\footnote{All simulations of diffraction patterns were performed using the computer code MOLTRANS.}
is shown in Fig.~\ref{Fig:ExpGeometr}(a). A coherent x-ray beam with a wavelength $\lambda=1.5\;\mathring{A}$ kinematically scatters from a 2D disordered sample of the total area $10\times10\;\mu\text{m}^{2}$.
Diffraction patterns are recorded on a 2D detector of the size $D=12\;\text{mm}$
(with the pixel size $p=10\;\mu\text{m}$), positioned in the far-field at the distance $L=2.5\;\text{m}$ from the sample.
The chosen scattering geometry allows to cover a $q$-range up to $0.1\;\text{nm}^{-1}$ (that corresponds the maximum scattering angle, $\theta_{\text{max}}=0.14^{\circ}$), and also corresponds to about four pixels per speckle in a diffraction pattern for a given sample size
\footnote{Sufficient resolution of the detector is important for a reliable determination of all contributions to the CCF
defined by Eqs.~(\ref{Cq1q2n_2}, \ref{Cq1q2n_T1}-\ref{Cq1q2n_T4}).}.

As a model of a disordered sample a 2D system composed of identical clusters, generally in random positions and orientations, is considered [Fig.~\ref{Fig:ExpGeometr}(b)]. We chose a centered pentagonal 2D cluster with a 5-fold rotational symmetry for our simulations [see the inset in Fig.~\ref{Fig:ExpGeometr}(b)].
This symmetry belongs to the forbidden motifs in the long-range crystalline matter but may exist on the short length scales in disordered systems\cite{Reichert}, and is, clearly, most intriguing for a detailed study.
Each cluster is composed of close-packed colloidal spheres with a radius of $100\;\text{nm}$ giving the total cluster size $d=600\;\text{nm}$.
The sample is oriented with its 2D plane perpendicular to the direction of the incident x-ray beam [see Fig.~\ref{Fig:ExpGeometr}(a)], and the 5-fold rotational axis of each pentagonal cluster in the sample is parallel to the direction of the incident x-ray beam.
The choice of parameters for our model was selected in analogy to the system studied  experimentally in  Ref.~\cite{Wochner1}.

In our simulations disordered samples are generated in the following way. Starting with an ordered 2D lattice of
clusters, positional disorder is achieved by a sequence of random movements of all clusters.  A single randomization step consists of Voronoi tessellation~\cite{Voronoj1} of the sample area, with each Voronoi cell containing only one cluster, and a subsequent displacement of each cluster by a random vector within the Voronoi cell. After $10-50$ steps there is no positional order in the system. This procedure is repeated several times to generate the sequence of uncorrelated systems.
Orientational disorder of clusters is implemented on the final step by applying a certain angular distribution of clusters to each of the generated samples. In Fig.~\ref{Fig:ExpGeometr}(b) one of the typical realizations of this procedure is presented.

\bigskip

(b) {\it Structural term}

\begin{figure*}[!htbp]
\centering
\includegraphics[width=0.55\textwidth]{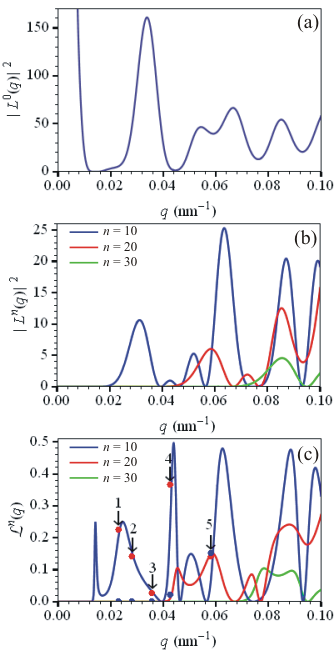}
\caption{\label{Fig:BesselFunc}(Color online)
The structural terms $|L^{n}(q)|^2$ [see Eqs.~(\ref{LocStr1}) and (\ref{AII.Pentagon})] calculated for a single centered pentagonal cluster as a function of $q$ for (a) $n=0$ and (b) $n=10$, 20 and 30. These terms are normalized by the value of the form-factor $|f(q)|^2$ of a sphere. (c) The normalized values $\mathcal{L}^{n}(q)=|L^{n}(q)/ L^{0}(q)|^2$ of the same Fourier components. Arrows indicate the $q$ values at which the CCFs presented in this paper were calculated (circles mark the values of the specific Fourier components at these positions).}
\end{figure*}

\bigskip
As discussed earlier, the contribution of a single cluster to different Fourier components of the CCF
is determined by the term $\mathcal{L}^{n}(q)$ in Eq.~(\ref{CnqDilute}) and strongly depends on its symmetry and structure.
In case of a centered pentagonal cluster considered in our simulations due to its 5-fold symmetry
only Fourier components with $n=10l$, where $l$ is an integer number, have a non-zero contribution (see Appendix A).
Importantly, the same term in  Eq.~(\ref{CnqDilute}) determines the $q$-dependence of Fourier components $C_{q}^{n}$ of the CCF.
In Fig.~\ref{Fig:BesselFunc} the values of $|L^{n}(q)|^{2}$ and  $|L^{0}(q)|^{2}$ normalized by the form-factor of a sphere $|f(q)|^2$ as well as their ratio $\mathcal{L}^{n}(q)=|L^{n}(q)/L^{0}(q)|^{2}$ as a function of $q$ are presented for $n=10, 20$ and $30$.
As a general rule, at low $q-$values, due to the properties of the Bessel functions, the strongest contribution to the CCF  comes from the Fourier component with the lowest $n$ ($n=10$ in our case). At higher $q-$values contribution of other components becomes comparable. Notice, that at some $q-$values the function $\mathcal{L}^{n}(q)$ contain sharp maxima corresponding to $q-$values where $|L^{0}(q)|^{2}$ has small values. In a dilute system where we can neglect inter-particle correlations the factor $NL^{0}(q)$ coincides with the structure factor $S(q)$ in SAXS.
It means that one can expect an enhancement of the structural term for small $S(q)$ values. Similar observations were made in experiment\cite{Wochner1}.
Examination of Fig.~\ref{Fig:BesselFunc}(c) also shows that small variations of $q-$values can result in large variations in the values of $\mathcal{L}^{n}(q)$ and consequently in $C_{q}^{n}$. This is also similar to results obtained in Ref.\cite{Wochner1}.
It can be also seen in  Fig.~\ref{Fig:BesselFunc}(c) that there are some regions in $q$ where the contributions of all components are close to zero. We may expect, that at these specific $q-$values the contribution of other terms in Eq.~(\ref{Cq1q2n_2}), that are responsible for inter-cluster correlations in dense systems, can become significant.

The values of the structural term $\mathcal{L}^{n}(q)$ at five selected values of $q$ [shown by arrows in  Fig.~\ref{Fig:BesselFunc}(c)] are
presented in Table~\ref{Table1}.
In the following we will compare these values with the values obtained for different systems by means of the XCCA.

\section{\label{Sec4} Large number of clusters in the system}

\begin{figure*}[!htbp]
\centering
\includegraphics[width=0.53\textwidth]{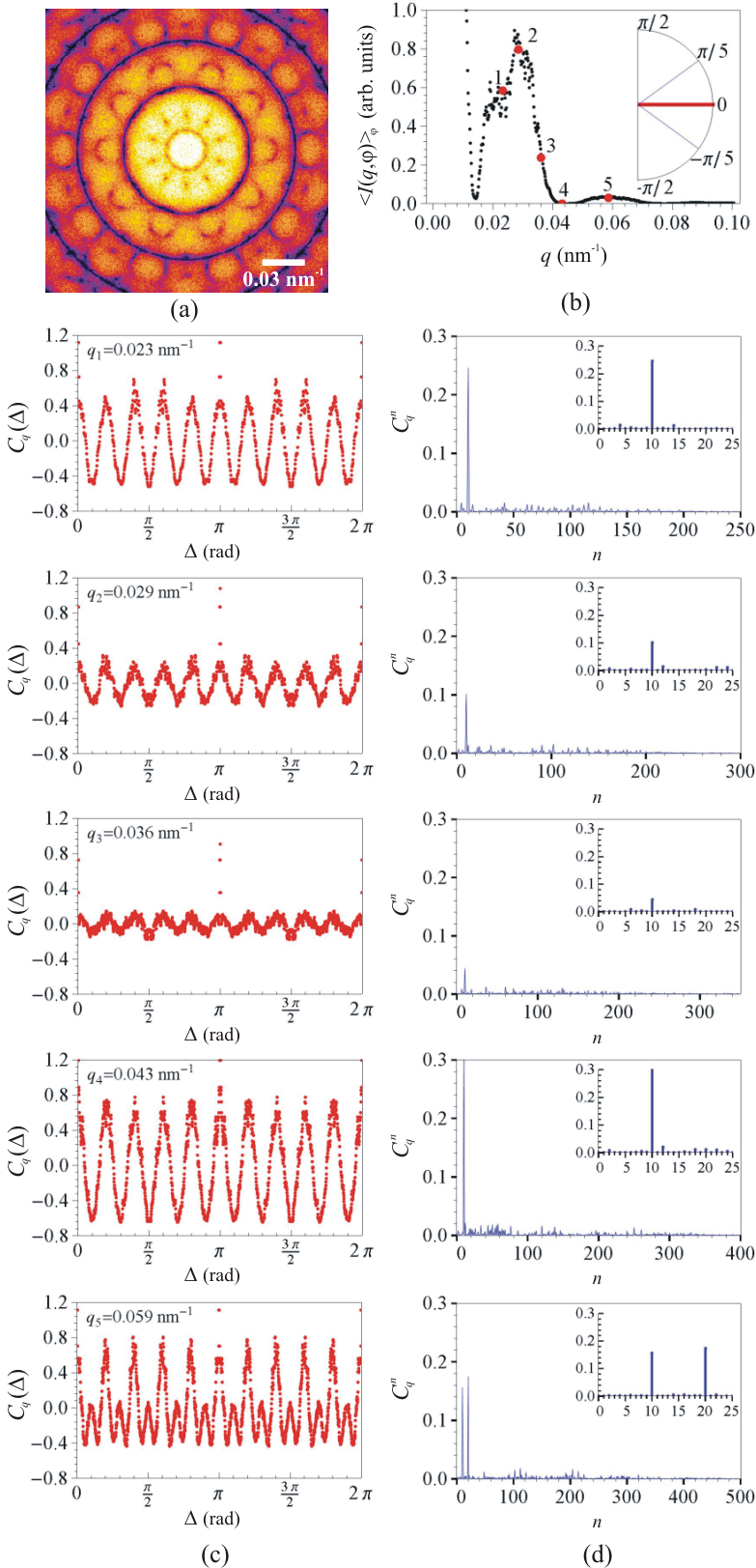}
\caption{\label{Fig:PentagonCase1}(Color online)
Cross-correlation analysis of the diffraction data for a completely oriented dense system ($\langle R \rangle/d\approx1.5$) consisting of 121 pentagonal clusters. (a) Diffraction pattern; (b) Angular averaged intensity $\langle I(q,\varphi)\rangle_{\varphi}$. (inset) Angular distribution of clusters in the sample. In this case all clusters have the same orientation; (c) CCFs $C_{q}(\Delta)$ calculated at the selected $q$ values, indicated in (b); (d) Fourier spectra $C_{q}^{n}$ of the CCFs for each selected $q$ value. (insets) Enlarged view of the first 25 Fourier components.}
\end{figure*}

\begin{figure*}[!htbp]
\centering
\includegraphics[width=0.55\textwidth]{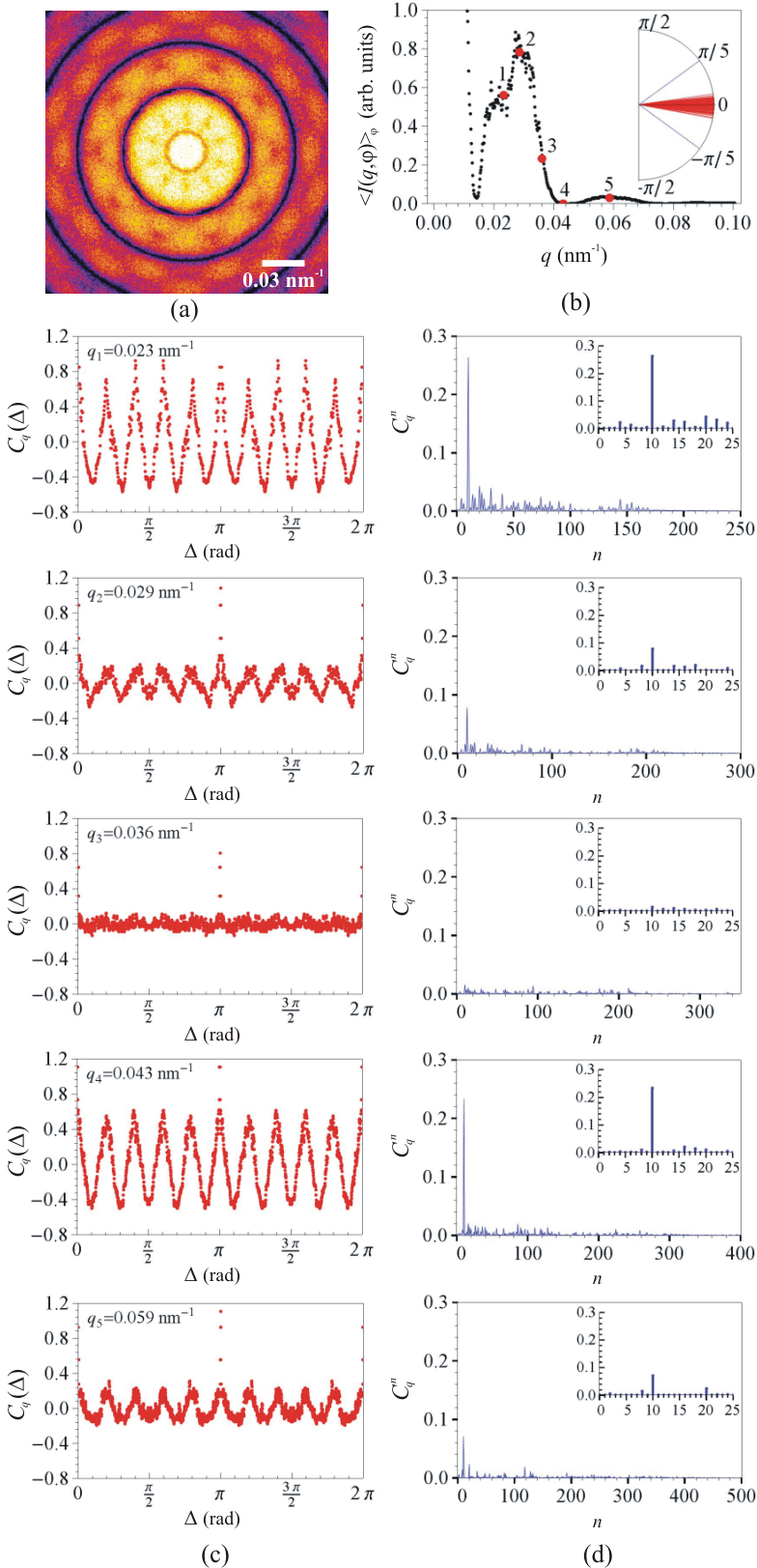}
\caption{\label{Fig:PentagonCase2}(Color online)
Cross-correlation analysis of the diffraction data (similar to Fig.~\ref{Fig:PentagonCase1}) for a system consisting of 121 pentagonal clusters described by the Gaussian distribution of cluster orientations with the standard deviation $\sigma_{\phi}=0.05\cdot 360^{\circ}/5=3.6^{\circ}$ and zero mean [see inset in (b)].}
\end{figure*}

\begin{figure*}[!htbp]
\centering
\includegraphics[width=0.55\textwidth]{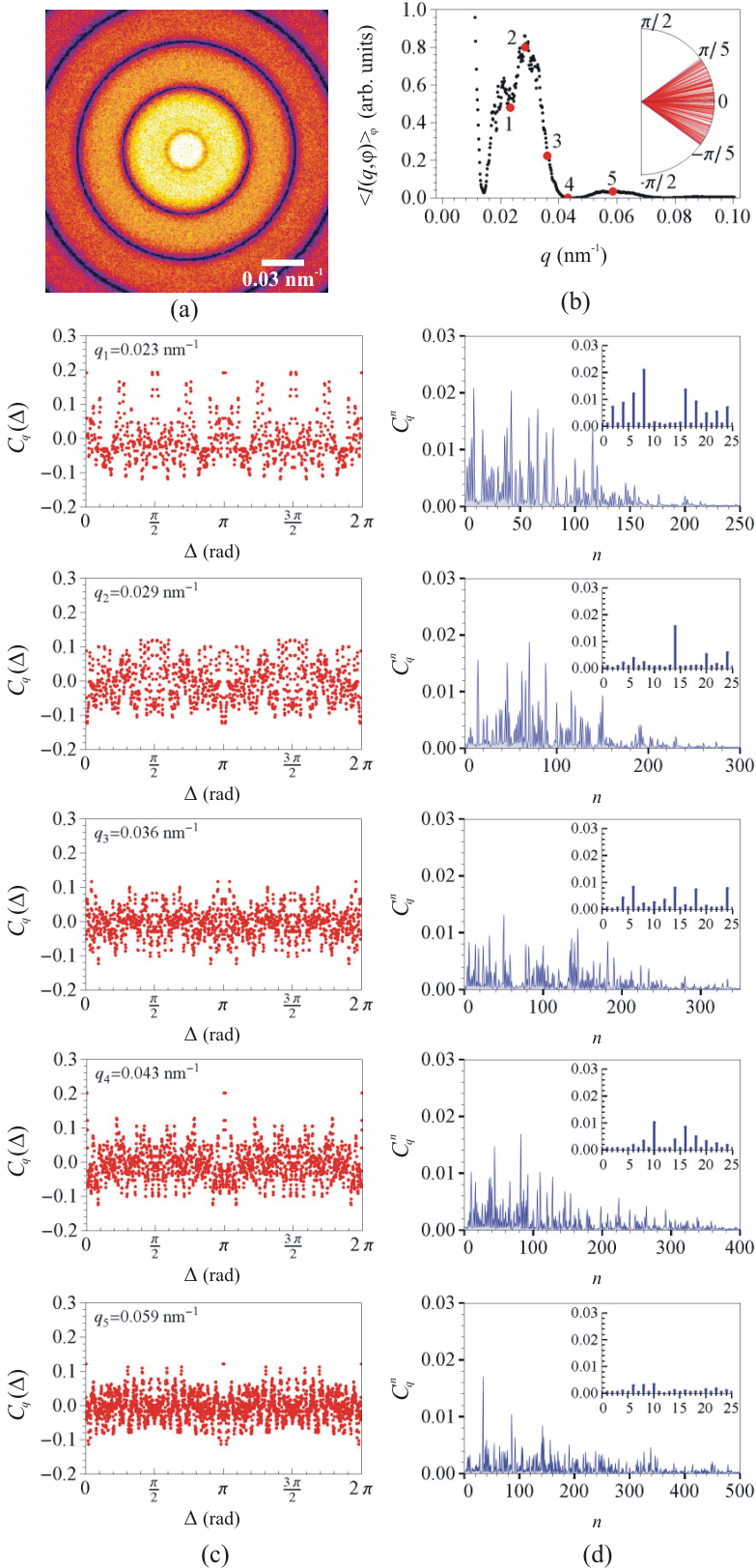}
\caption{\label{Fig:PentagonCase3}(Color online)
Cross-correlation analysis of the diffraction data (similar to Figs.~\ref{Fig:PentagonCase1} and \ref{Fig:PentagonCase2})
for a system consisting of 121 pentagonal clusters described by the uniform distribution of cluster orientations in the angular range $(-\pi/5, \pi/5)$ [see inset in (b)].}
\end{figure*}

In this section we consider model systems with different characteristics of orientational disorder. We are specifically interested here in analyzing the case of a dense system, contrary to the analysis of the dilute systems performed in our previous paper\cite{Altarelli}. In the following we consider three systems consisting of 121 pentagonal clusters
\footnote{For the given sample area ($10\times10\;\mu\text{m}^{2}$) and cluster size ($d=600\;\text{nm}$) one can simulate a close-packed system composed of more than 200 clusters. However, such a high density of clusters is not convenient for modeling of a disordered system since it may induce some positional ordering in the system. To achieve better positional disorder and at the same time to keep a relatively high density of clusters we consider here the systems composed of 121 clusters.}
with the same positions of clusters in all three systems but with different orientational disorder in each system [see
Fig.~\ref{Fig:ExpGeometr}(b)].
These three systems are: a completely oriented system of clusters (Fig.~\ref{Fig:PentagonCase1}), a partially oriented  system described by the Gaussian distribution of orientations with the standard deviation $\sigma_{\phi}=0.05\cdot360^{\circ}/5=3.6^{\circ}$ and zero mean
(Fig.~\ref{Fig:PentagonCase2}) and one with a uniform distribution of orientations in the angular range
\footnote{Since a pentagonal cluster has 5-fold rotational symmetry, any rotation of the cluster about its 5-fold rotational axis can be associated with the angular orientation in the interval $(-\pi/5,\pi/5)$. In other words, any rotation $-\infty<\phi_{k}<\infty$ can be expressed as an orientation $\phi_{k}^{0}=\phi_{k} \text{ mod } \pi/5$, which is defined in the interval $-\pi/5\leq\phi_{k}^{0}<\pi/5$. That is why, it is sufficient for a pentagonal cluster to consider the uniform distribution of its angular orientations in the interval $(-\pi/5,\pi/5)$. However, one should be careful in simulations where clusters have a nonuniform distribution of orientations. For example, in the case of a broad Gaussian distribution of orientations of pentagonal clusters, where a possible rotation of the cluster may fall outside the interval $-\pi/5\leq\phi_{k}<\pi/5$, one should rather deal with a wrapped distribution of orientations\cite{CircStat1, CircStat2}, which falls into the scope of directional statistics. In this paper we choose only such sets of orientations of pentagonal clusters (both for the uniform and for the Gaussian distributions), where all the angles satisfy the condition $-\pi/5\leq\phi_{k}<\pi/5$.}
$(-\pi/5,\pi/5)$ (Fig.~\ref{Fig:PentagonCase3}). The relative separation between the clusters in all three cases is characterized by the ratio $\langle R\rangle /d\approx1.5$, where $\langle R\rangle$ is the average distance between the clusters estimated from $\langle R\rangle\sim1/\sqrt{n_{0}}$, and $n_{0}$ is the number of clusters per unit area (number density).

Diffraction patterns for each case are presented in Figs.~\ref{Fig:PentagonCase1}(a), \ref{Fig:PentagonCase2}(a) and \ref{Fig:PentagonCase3}(a). The most prominent features for all diffraction patterns are concentric rings
appearing due to the contribution of the form-factor $f(q)$ of a  single colloidal sphere to the scattered intensity.
The speckles superimposed on the diffraction patterns originate from the interference of x-rays coherently scattered on the spatially disordered clusters, distributed within a finite sample area.  The corresponding angular averaged intensities $\langle I(q,\varphi)\rangle_{\varphi}$ for each diffraction pattern, as well as the angular distribution of orientations of clusters in each system are presented in Figs.~\ref{Fig:PentagonCase1}(b), \ref{Fig:PentagonCase2}(b) and \ref{Fig:PentagonCase3}(b). The following momentum transfer $q-$values: $q_{1}=0.023\;\text{nm}^{-1}$, $q_{2}=0.029\;\text{nm}^{-1}$, $q_{3}=0.036\;\text{nm}^{-1}$, $q_{4}=0.043\;\text{nm}^{-1}$ and $q_{5}=0.059\;\text{nm}^{-1}$ were considered for the detailed analysis in each case [shown in Figs.~\ref{Fig:PentagonCase1}(b), \ref{Fig:PentagonCase2}(b) and \ref{Fig:PentagonCase3}(b) by red circles]. The CCFs $C_q(\Delta)$ as well as corresponding Fourier components  $C_{q}^{n}$ were evaluated using Eqs.~(\ref{Eq:Cq1q2_0}, 2b) for each system at these selected $q-$values and are presented in Figs.~\ref{Fig:PentagonCase1}(c,d), \ref{Fig:PentagonCase2}(c,d) and \ref{Fig:PentagonCase3}(c,d).

Comparison of all three cases shows that similar to a dilute system the dominant Fourier components ($n=10$ and 20 in our case) are well defined for more oriented systems (Fig.~\ref{Fig:PentagonCase1} and Fig.~\ref{Fig:PentagonCase2}).
It is also well seen that for a Gaussian distribution of orientations (Fig.~\ref{Fig:PentagonCase2}) at higher $q-$values ($q=q_5$ in this case) the magnitudes of the Fourier components with $n=10$ and $n=20$ are significantly reduced in comparison to a completely oriented system (Fig.~\ref{Fig:PentagonCase1}). This decay can be attributed to the exponential factor present in the statistical term in Eqs.~(\ref{GausDistr},\ref{2.26}).

Our analysis of the first two systems with the high degree of orientational order shows that the symmetry of clusters composing the system can be well determined by means of the XCCA analysis. However, the absolute values of Fourier components can be strongly affected by the inter-particle correlations (see Tables~\ref{Table1},~\ref{Table2}). Comparison of the absolute values of the Fourier components $C_{q}^{n}$ at $n=10, 20$  for the completely oriented system considered here shows that they may fluctuate around the value of the structural term $\mathcal{L}^{n}(q)$ for about few tens of percent at different $q-$values (see Table~\ref{Table1}).
Results of similar calculations for a dense system with the Gaussian distribution of orientations are presented in Table~\ref{Table2}. They are compared with the averaged Fourier components of CCFs defined in Eq.~(\ref{2.26}). They show similar fluctuations of the values of $C_{q}^{n}$ due to inter-particle correlations in the system and statistical fluctuations of orientational distribution of clusters (see next section for a detailed discussion).

Our simulations for the systems with  different positional disorder have shown that $C_{q}^{n}$ also depend on the particular realization of the system.
Such behaviour of $C_{q}^{n}$ can be explained by the specific constructive and destructive interference of the scattered x-rays in each particular system. Due to the interference term ($S^{n}_{2}+S^{n}_{3}$) in Eq.~(\ref{Cq1q2n_2}) the inter-particle contribution can increase or decrease the contribution of the structural term  $S^{n}_{1}$ in a dense system. 
Note, that the term  $S^{n}_{4}$ is always non-negative and can only increase the value of $C_{q}^{n}$.
That brings us to the conclusion that determination of the absolute values of the Fourier coefficients $C_{q}^{n}$, or $|I^{n}(q)|$ that is necessary for the direct structural determination of the clusters by phase retrieval \cite{Saldin1} can be strongly affected in close packed systems.

Contrary to these two cases, the system with the uniform distribution of orientations shows quite different behavior. As a result of the complete orientational disorder, the CCFs shown in Fig.~\ref{Fig:PentagonCase3}(c) do not represent a simple $\sim \cos(n\Delta)$ form but have rather complicated angular dependence that can be also well observed in the corresponding Fourier spectra in Fig.~\ref{Fig:PentagonCase3}(d). For example, the component with $n=8$ (and some other higher values of $n$) dominates at $q=q_{1}$ contrary to the previous more ordered systems, where the component with $n=10$ was strongly dominant at that $q-$value. It is well seen in Fig.~\ref{Fig:PentagonCase3}(d) that in this case there are many Fourier components with comparable values in a wide range of the spectrum. We should note also here that the absolute values of all Fourier components in this case are about one order of magnitude lower then in the case of the ordered systems [see Figs.~\ref{Fig:PentagonCase1}(d) and \ref{Fig:PentagonCase2}(d)]. This general behavior of disordered systems can be also explained by examining the contribution of two terms in Eq.~(\ref{Inq}). In the system with the uniform distribution of angular orientations the absolute value of the first term in Eq.~(\ref{Inq}) is significantly reduced due to the asymptotic behavior of the statistical term $A^2_n$ given in Eq.~(\ref{UnifDistr}). For a large number of orientations it drops as $1/N$ and becomes much smaller then the second term in Eq.~(\ref{Inq}) that is determined mostly by the inter-particle correlations. In this case inter-particle correlations give the major contribution to CCFs and determination of the structure or symmetry of the clusters composing the system becomes practically impossible.

The results of our simulations for the 2D samples with different degree of orientational disorder presented in this section clearly indicate, that the possibility to extract the information on the symmetry of individual clusters in the systems with a large number of clusters by the cross-correlation analysis is strongly related to the degree of orientational disorder in the system. The preferred situation is when the system is partially aligned, while correlation analysis is practically impossible in a completely disordered dense system.


\section{Ensemble-dependent fluctuations of correlations}

In this section we will consider in more detail the statistical behavior of
the CCFs as a function of orientational disorder in a system with a small number of possible orientations. As was discussed above, in a dilute disordered system composed of a certain number of clusters the statistical term $A_{n}^2$ in the CCF is fluctuating around an average value $\langle A_{n}^{2} \rangle$ [see Eqs.~(\ref{GausDistr}) and (\ref{UnifDistr})], determined by the statistics (Gaussian, uniform, {\it etc.}) and number $N$ of angular orientations of clusters.
Contrary to the previous section where we analyzed systems with a large number of clusters ($N=121$), here we consider systems with much smaller number of clusters for which statistical fluctuations of the scattered intensity become considerably larger.

\begin{figure*}[!htbp]
\centering
\includegraphics[width=1.0\textwidth]{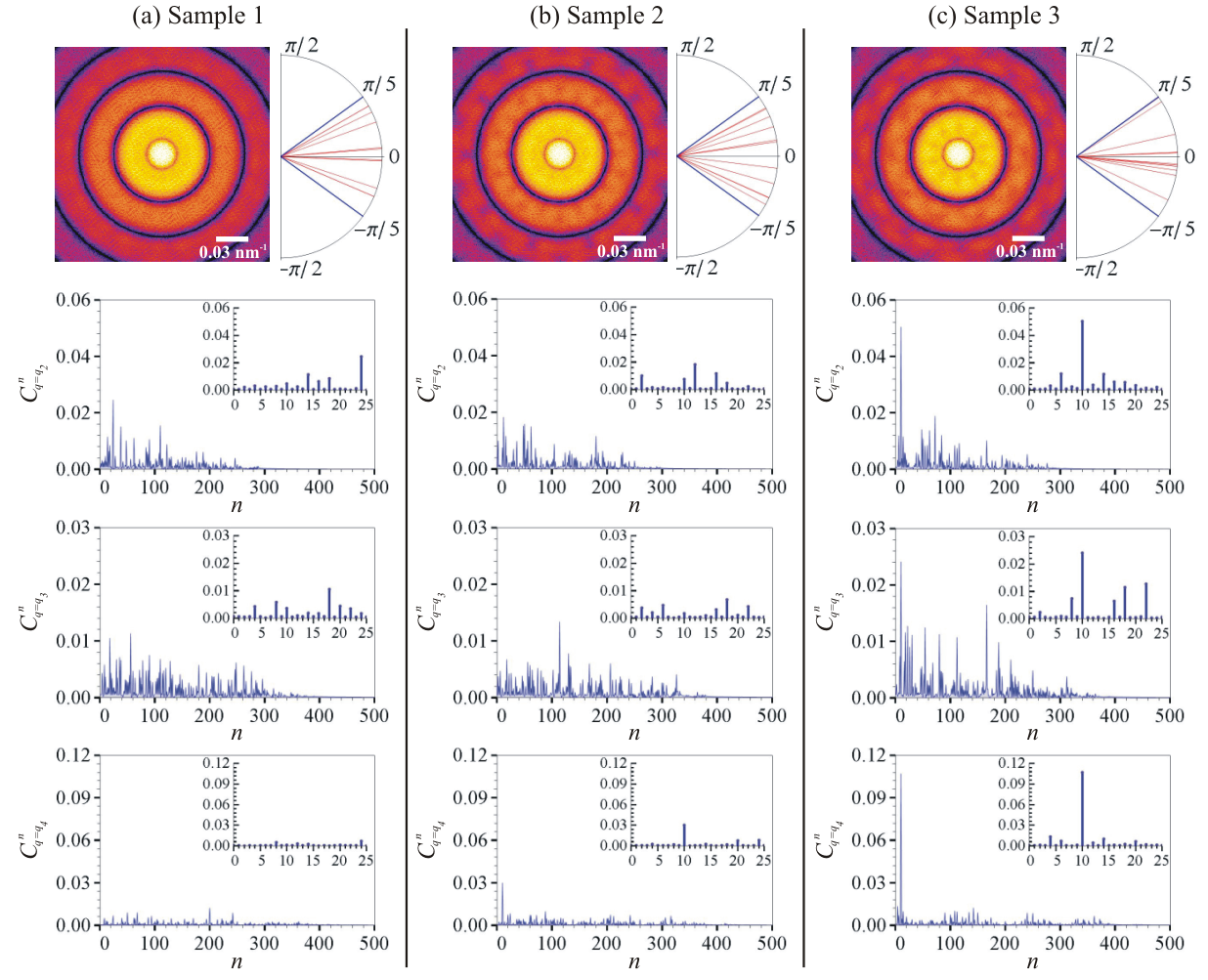}
\caption{\label{Fig:EnsembleCompare}(Color online)
Diffraction patterns and Fourier components of the CCFs calculated for three disordered systems, containing 11 clusters in different, uniformly distributed orientations. Angular distribution of clusters in each sample is given in the angular diagrams. Fourier spectra were calculated for each system at three $q$ values, $q_{2}= 0.029\;\text{nm}^{-1}$,
$q_{3}= 0.036\;\text{nm}^{-1}$, and $q_{4}= 0.043\;\text{nm}^{-1}$.}
\end{figure*}

We analyze here three samples, each composed of $N=11$ clusters. The positions of clusters are the same in each sample, and their relative separation is given by the ratio $\langle R\rangle /d\sim5.0$.
We consider for each sample a uniform distribution of 11 orientations of clusters within the angular range $(-\pi/5,\pi/5)$ [see Fig.~\ref{Fig:EnsembleCompare}].

The results of the Fourier analysis of the CCFs for these three systems at the selected values of $q$ are shown in Fig.~\ref{Fig:EnsembleCompare} (note that a different vertical scale for the Fourier spectra was used at different $q$ values). Comparing the Fourier spectra calculated at the same $q$ value for these systems one can see, that they significantly vary only due to a different orientational distribution of clusters from one sample to another.
In the case of dilute systems, fluctuations of the values of Fourier components are defined by the statistical distribution of orientations of clusters [see Eqs.~(\ref{CnqDilute}-\ref{UnifDistr})].
In the systems considered here, these fluctuations are
defined by the interplay of two factors, the distribution of  cluster orientations [term $S^{n}_{1}$ in expansion (\ref{Cq1q2n_2})], and their relative spatial positions (terms $S^{n}_{2}$, $S^{n}_{3}$ and $S^{n}_{4}$).
This leads in some cases to a large contribution of $S^{n}_{1}$ , sufficient to dominate over the other terms in Eq.~(\ref{Cq1q2n_2}) [see, for example, the results for the Sample 3 in Fig.~\ref{Fig:EnsembleCompare}(c)], and in some cases to a smaller contribution, allowing the inter-cluster correlations to dominate in the Fourier component $C^{n}_{q}$ [see, for example, the results for the Sample 1 in Fig.~\ref{Fig:EnsembleCompare}(a)]. Therefore, even if the number of possible orientations of clusters in a system is small, it may be not possible to extract the information on local structure of such a system from a single diffraction pattern.
The statistical behavior of CCFs discussed here could give a possible explanation to experimental observations in Ref.\cite{Wochner1}, where diffraction patterns taken at different times show different types of angular correlations.

As demonstrated here in a system with a small number of randomly oriented clusters reliable information on the local structure and symmetry of the clusters can not be obtained by the analysis of a single diffraction pattern due to statistical fluctuations of intensity. We discuss in the next section how this problem can be solved in certain cases by averaging of CCFs.


\section{Effects of orientational disorder on the averaged CCF}

In our previous discussion we presented the results of the XCCA applied to the analysis of the
diffraction data obtained for a single realization of a disordered system.
We showed that statistical fluctuations of CCFs from one realization to another is a strong
limitation for the determination of reliable information on cluster symmetry and structure.
To overcome this difficulty it was proposed, first by Kam\cite{Kam, Kam1},
to analyze CCFs averaged over an ensemble of diffraction patterns, instead of analyzing CCFs calculated for a single diffraction pattern (see also recent publications \cite{Saldin, Saldin1, Saldin2}).
Below we will investigate conditions when this approach is valid.

We will consider here the CCF $\langle C_{q}(\Delta) \rangle_{M}$ averaged over a sufficiently large number $M$ of diffraction patterns,
\begin{equation}
\langle C_{q}(\Delta) \rangle_{M} = 1/M \sum\limits_{m=1}^{M} C^{m}_{q}(\Delta),
\label{AverCorr1}
\end{equation}
where $C^{m}_{q}(\Delta)$ is the CCF obtained for the $m\text{-th}$ diffraction pattern [Eq.~(\ref{Eq:Cq1q2_0})]. 
Performing the Fourier transform of both parts of Eq.~(\ref{AverCorr1}) and taking into account that Fourier transform is a linear operator we obtain for the Fourier components of the averaged CCFs
\begin{equation}
\langle C^{n}_{q} \rangle_{M} =1/M \sum\limits_{m=1}^M \{C^{n}_{q}\}^{m},
\label{AverCorr2}
\end{equation}
where $\{C^{n}_{q}\}^{m}$ are the $n-$th Fourier components of the $m\text{-th}$ CCF $C^{m}_{q}(\Delta)$. Eq.~(\ref{AverCorr2}) means, that the $n-$th Fourier component of the averaged CCF is an average of the corresponding Fourier components determined for each diffraction pattern.
Taking into account expression (\ref{Cq1q2n_2}) for each individual Fourier component of the CCF
we obtain
\begin{equation}
\langle C^{n}_{q} \rangle_{M} =\left\langle \left\vert \frac{I^{n}(q)}{I^{0}(q)} \right\vert^2 \right\rangle_{M}=
\left\langle\frac{ S^{n}_{1}}{\vert I^{0}(q)\vert^2} \right\rangle_{M}+
\left\langle \frac{(S^{n}_{2}+S^{n}_{3})}{\vert I^{0}(q)\vert^2}\right\rangle_{M}+
\left\langle \frac{S^{n}_{4}}{\vert I^{0}(q)\vert^2}\right\rangle_{M},
\label{AverCorr2a}
\end{equation}
where $S^{n}_{j}$ are defined by Eqs.~(\ref{Cq1q2n_T1}-\ref{Cq1q2n_T4}) and $\vert I^{0}(q)\vert^2$ by Eq.~(\ref{Cq1q2n_I0}).
Below we show by direct simulations that for a sufficiently large number of diffraction patterns $M$
the ensemble averaged term $\langle (S^{n}_{2}+S^{n}_{3})/\vert I^{0}(q)\vert^2\rangle_{M}$ asymptotically approaches zero and we have for the Fourier components of the averaged CCF
\begin{equation}
\langle C^{n}_{q} \rangle_{M} =\langle C^{n}_{q\;(clust)}\rangle_{M}+\langle C^{n}_{q\;(int-clust)}\rangle_{M},
\label{AverCorr2b}
\end{equation}
where
\begin{subequations}
\begin{eqnarray}
&&\langle C^{n}_{q\;(clust)}\rangle_{M}=\langle S^{n}_{1} / \vert I^{0}(q)\vert^2 \rangle_{M},\label{Eq:ClustAver}\\
&&\langle C^{n}_{q\;(int-clust)}\rangle_{M}=\langle S^{n}_{4} / \vert I^{0}(q)\vert^2 \rangle_{M}.\label{Eq:IntClustAver}
\end{eqnarray}
\end{subequations}
This result indicates that in a general case for a sufficiently large number of diffraction patterns the
Fourier components of the averaged CCF can be represented as an additive sum of two positive valued contributions. The first one
is given by the structure of individual clusters and the second one is determined by the inter-cluster
contribution.

In the case of a dilute disordered system, when the inter-cluster contribution $\langle C^{n}_{q\;(int-clust)}\rangle_{M}$
in Eq.~(\ref{AverCorr2b}) can be neglected, we obtain
\begin{equation}
\langle C^{n}_{q} \rangle_{M}=\langle C^{n}_{q\;(clust)}\rangle_{M} = \mathcal{L}^{n}(q)\cdot \langle A^{2}_{n} \rangle_M,
\label{AverCorr3}
\end{equation}
where $\langle A^{2}_{n} \rangle_M$ is an ensemble averaged square amplitude of the random phasor sum (\ref{2.24})
\begin{equation}
\langle A^{2}_{n} \rangle_{M}=1/M\sum\limits_{m=1}^{M}  \left\{A^{2}_{n}\right\}^{m}.
\label{AverCorr4}
\end{equation}

For a sufficiently large number $M$ of diffraction patterns, the value of $\langle A^{2}_{n} \rangle_{M}$ approaches its statistical limit $\langle A_{n}^{2} \rangle$ [see Eqs.~(\ref{GausDistr}) and (\ref{UnifDistr})].
This result immediately explains the advantage of calculation of the average CCFs
in dilute systems as soon as it converges to a finite not fluctuating value. This opens the opportunity to determine the structural term $\mathcal{L}^{n}(q)$
for systems with orientational disorder. At the same time our analysis shows that the favorable situation for such averaged analysis are dilute systems with a small number of orientations. If the number $N$ of orientations increases, contribution of this averaged statistical term $\langle A_{n}^{2} \rangle_M$ decreases [for example, as $\sim1/N$ for a uniform distribution of orientations according to Eq.~(\ref{UnifDistr})], and the structural contribution
$\langle C^{n}_{q\;(clust)}\rangle_{M}$ can become smaller than the inter-cluster contribution $\langle C^{n}_{q\;(int-clust)}\rangle_{M}$. In Appendix~\ref{AppC} we give a detailed derivation for the statistical estimate of the inter-cluster contribution to $\langle C^{n}_{q} \rangle_{M}$ [Eq.~(\ref{AverCorr2b})].

One important question that is relevant to the discussion in this chapter is how many diffraction patterns $M$ will be sufficient for calculation of the average $\langle A^{2}_{n} \rangle_M$ in order to reach reliably its statistical limit $\langle A^{2}_{n} \rangle$. In other words, what is an averaging accuracy $\varepsilon$ for a number $M$ of diffraction patterns. This question can be addressed by establishing a confidence interval for $\langle A^{2}_{n} \rangle_M$.
Our analysis has shown (see Appendix~\ref{AppB} for details) that in the case of a uniform distribution of cluster orientations and averaging accuracy of $10\%$ the probability that $\langle A^{2} \rangle_{M}$ lies in the interval from $0.9\langle A^{2}\rangle$ to $1.1\langle A^{2} \rangle$ is more than $90\%$ for $M=10^3$, and more than $99\%$ for $M=10^4$ diffraction patterns.

\begin{figure*}[!htbp]
\centering
\includegraphics[width=1.0\textwidth]{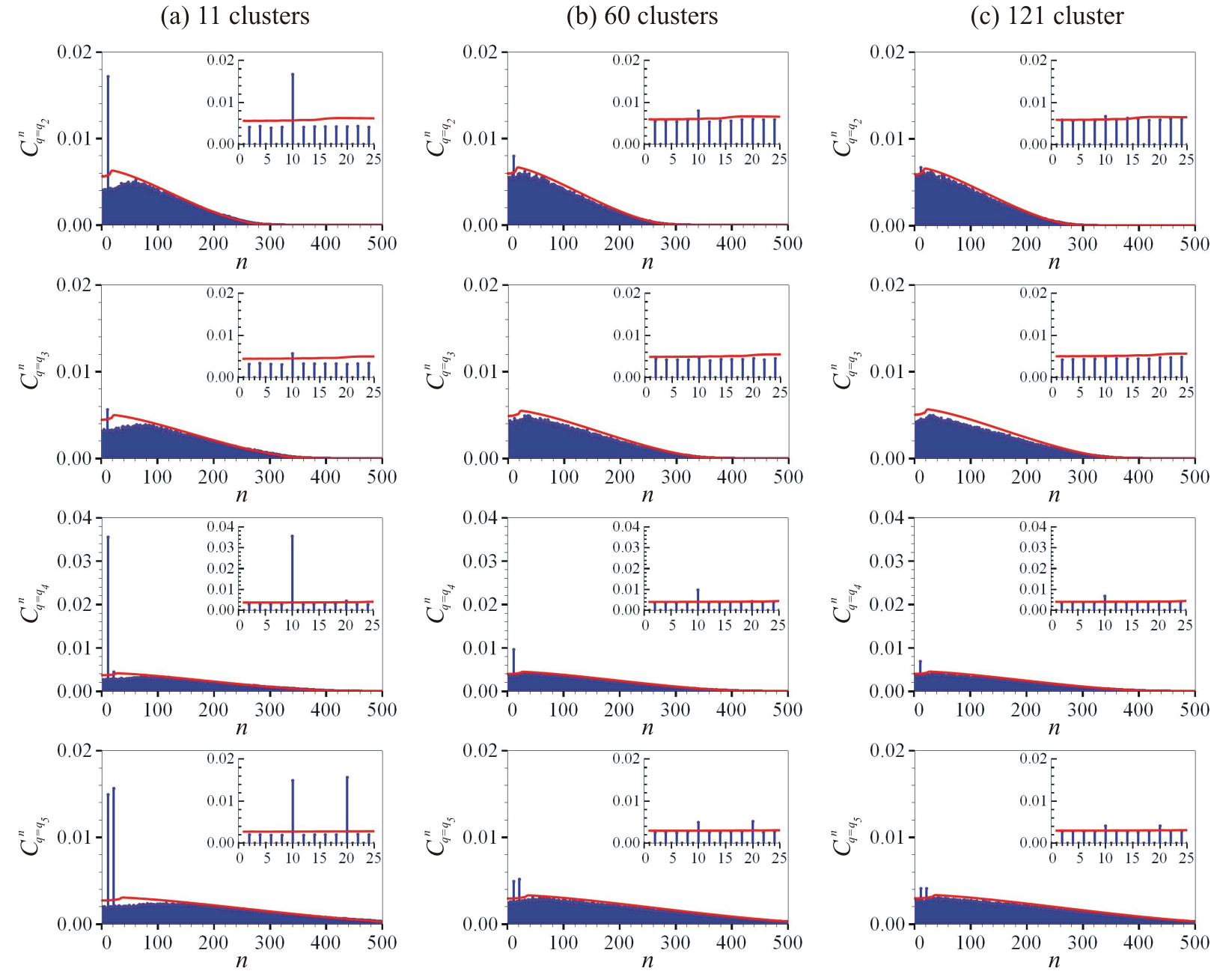}
\caption{\label{Fig:CorrAver}(Color online)
Fourier spectra of the CCFs $\langle C_{q}(\Delta) \rangle_{M}$, averaged over $M=1000$ diffraction patterns. Three disordered systems composed of (a) 11, (b) 60 and (c) 121 clusters with the uniform distribution of angular orientations are considered. Fourier spectra were calculated for each system at four $q$ values, $q_{2}= 0.029\;\text{nm}^{-1}$,
$q_{3}= 0.036\;\text{nm}^{-1}$, $q_{4}= 0.043\;\text{nm}^{-1}$ and $q_{5}= 0.059\;\text{nm}^{-1}$. Red solid line in each Fourier spectrum defines the magnitude of the inter-cluster contribution estimated using Eq.~(\ref{C3}).}
\end{figure*}

We demonstrate these results by analyzing the ensemble averaged CCFs for three systems consisting of (a) 11, (b) 60 and (c) 121 clusters with the relative distances (a) $\langle R\rangle /d\sim5.0$, (b) $\langle R\rangle /d\sim2.2$ and (c) $\langle R\rangle /d\approx1.5$ respectively. For each realization of a system in the ensemble different set of angles and cluster positions were considered. In  Fig.~\ref{Fig:CorrAver} the results of the Fourier analysis of the CCFs averaged over $M=1000$ diffraction patterns are presented. According to our previous analysis, for this number of diffraction patterns the term $\langle C^{n}_{q\;(clust)}\rangle_{M}$ in Eq.~(\ref{AverCorr2b}) with the probability more then $90\%$ lies in the interval $\langle C^{n}_{q\;(clust)}\rangle\pm 10\%$.

Comparison of the averaged Fourier spectra for the system containing 11 clusters [Fig.~\ref{Fig:CorrAver}(a)], with the ones calculated for different single realizations of the same system [Fig.~\ref{Fig:EnsembleCompare}] shows a strong enhancement of the contrast of the contribution from the internal structure of the pentagonal cluster, as compared to the inter-cluster contribution. 
At the same time, comparison of the magnitudes of the same Fourier components with the ones obtained in 
a dilute limit for a system with the uniform distribution of cluster orientations [see Eq.~(\ref{AverCorr3}) 
with $M \to \infty$] reveals significant deviations [see Table~\ref{Table3}]. We attribute this effect to the presence of the inter-cluster contribution in the latter case.

Comparison of the results presented in Fig.~\ref{Fig:CorrAver} for three different samples shows, that as soon as the number of clusters in a dense system increases, the Fourier components $\langle C^{n}_{q\;(clust)}\rangle_{M}$ related to the local structure decrease following the $1/N$ dependence [see Eq.~(\ref{UnifDistr})].
It is well seen in Figs.~\ref{Fig:CorrAver}(b) and \ref{Fig:CorrAver}(c) that for a system with the number of clusters $N=60$ the Fourier components with $n=10$ and $20$ are already hardly resolved over the inter-cluster contribution, and for the system with $N=121$ clusters their contribution is not resolved over the level of inter-cluster contribution.
Therefore, if the number of possible orientations of LSs in a dense system is sufficiently large, as in Fig.~\ref{Fig:CorrAver}(c), calculation of the average CCFs may not provide information on the local structure in a disordered system.

The contribution of the inter-cluster correlations $\langle C^{n}_{q\;(int-clust)}\rangle_{M}$ to the averaged CCF is well seen in an entire Fourier spectrum
for all three systems presented in Fig.~\ref{Fig:CorrAver}.
In Appendix~\ref{AppC} an asymptotic estimate of the inter-cluster contribution in $\langle C^{n}_{q} \rangle_{M}$ [Eq.~(\ref{AverCorr2b})]
is obtained (see solid red line in Fig.~\ref{Fig:CorrAver}).
As one can see in Fig.~\ref{Fig:CorrAver}, our estimate of the inter-cluster contribution quite accurately reproduces the results of the direct calculations for all systems considered here.
Our analysis has shown that for a sufficiently large number $M$ of diffraction patterns the magnitude of the averaged inter-cluster contribution $\langle C^{n}_{q\;(int-clust)}\rangle_{M}$ depends
on the shape and size of the sample, as well as on the size of particles, and rather weakly on their number $N$.

\begin{figure*}[!htbp]
\centering
\includegraphics[width=1.0\textwidth]{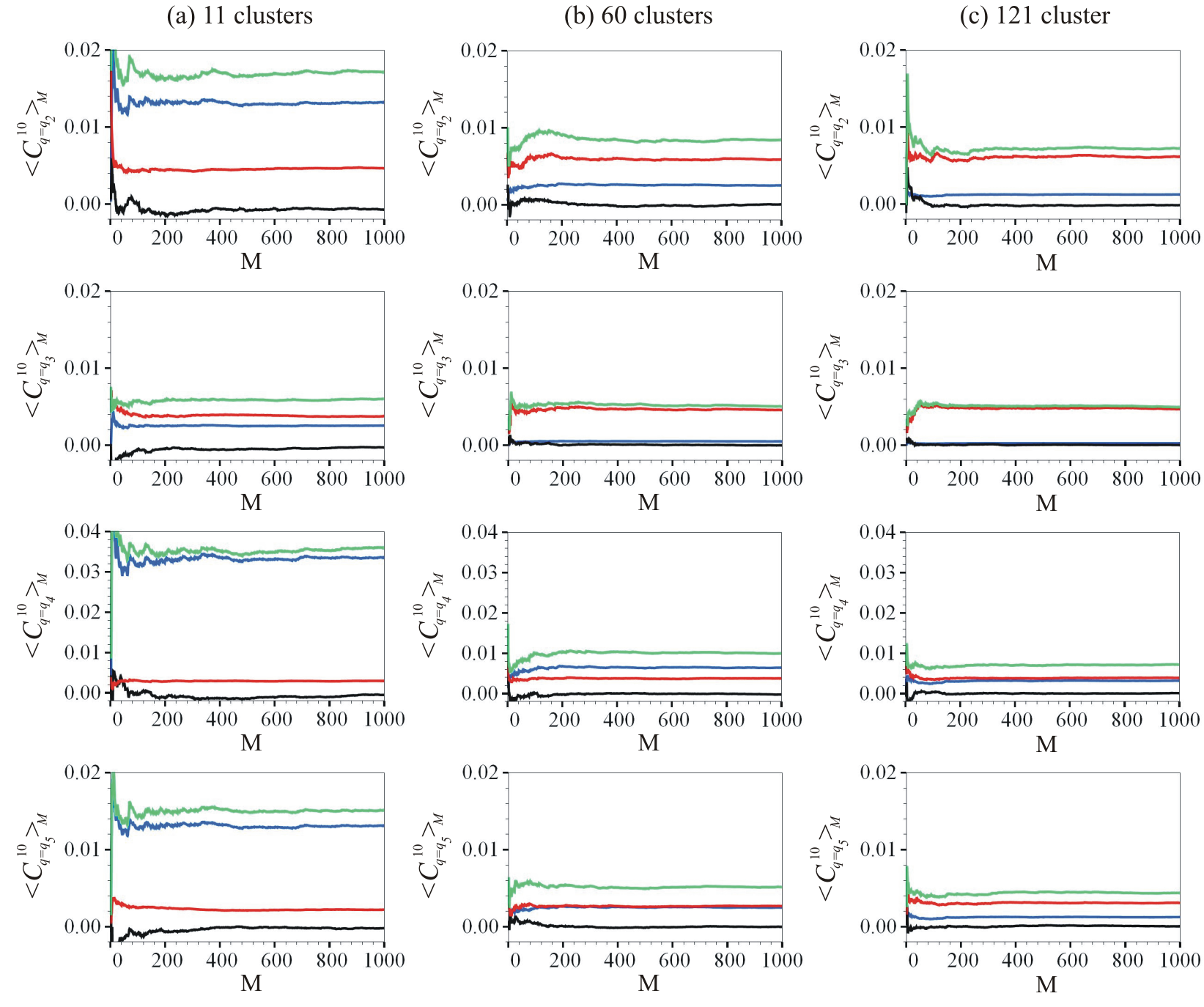}
\caption{\label{Fig:CorrAverEvolution}(Color online)
(Color online) The evolution of different terms in the expansion of the Fourier component $\langle C_{q}^{n}\rangle_{M}$ $(n=10)$ [Eq.~(\ref{AverCorr2a})] as a function of the number $M$ of diffraction patterns used in the averaging. The results are presented for the same $q$ values as in Fig.~\ref{Fig:CorrAver}. The following contributions are shown: $\langle S^{n}_{1}\rangle_{M}/\langle \vert I^{0}(q)\vert^2 \rangle_{M}$ (blue curve),  $\langle S^{n}_{2}+S^{n}_{3}\rangle_{M}/\langle \vert I^{0}(q)\vert^2 \rangle_{M}$ (black curve), $\langle S^{n}_{4}\rangle_{M}/\langle \vert I^{0}(q)\vert^2 \rangle_{M}$ (red curve), and the sum of all terms $\langle C_{q}^{n}\rangle_{M}$ (green curve). Three disordered systems composed of (a) 11, (b) 60 and (c) 121 clusters with the uniform distribution of angular orientations are considered.}
\end{figure*}

The evolution of different terms in the expansion of the averaged Fourier component $\langle C_{q}^{n}\rangle_{M}$ [Eq.~(\ref{AverCorr2a})] as a function of the number $M$ of diffraction patterns is presented in Fig.~\ref{Fig:CorrAverEvolution}. The results are shown for $C_{q}^{n}$ with $n=10$ for the same systems and $q$ values as in Fig.~\ref{Fig:CorrAver}. These results demonstrate the convergence of different terms to their average values. In particular, as one can see from Fig.~\ref{Fig:CorrAverEvolution}, the averaged term 
$\langle (S^{n}_{2}+S^{n}_{3})/\vert I^{0}(q)\vert^2 \rangle_{M}$ in Eq.~(\ref{AverCorr2a})  asymptotically reaches zero and can be neglected after averaging over a few hundred diffraction patterns. For a system with the number of particles $N=11$ the structural term strongly dominates over the inter-cluster contribution at most of the $q$ values (it is only lower for $q=q_3$). Contrary to that case, for the systems with the number of clusters $N=60$ and $N=121$ the inter-cluster contribution term
is larger than the term corresponding to the structure of individual clusters
for most of the $q-$values. That is the reason why the structural contribution could be hardly resolved for a system with $N=121$ clusters in Fig.~\ref{Fig:CorrAver}(c).

Our results clearly show that, in the general case, the ensemble averaged CCF $\langle C_{q}(\Delta) \rangle_{M}$
contains both the contribution from the individual clusters and
the inter-cluster contribution [see Eq.~(\ref{AverCorr2b})]. Therefore, the ensemble averaged CCF $\langle C_{q}(\Delta) \rangle_{M}$ determined from a system of coherently illuminated clusters, in general, can not be considered as a function of only `single-particle quantities' as it was stated in Refs.~\cite{Saldin1, Saldin3}.
For a large number $N$ of cluster orientations in the system the inter-cluster contribution
$\langle C^{n}_{q\;(int-clust)}\rangle_{M}$ can become dominant over the structural contribution
$\langle C^{n}_{q\;(clust)}\rangle_{M}$.
In the next section we consider a partially coherent scattering from disordered systems as one of the possible ways to suppress the inter-particle contribution to the CCF.


\section{Effects of partial coherence}

In the previous sections we analyzed diffraction data obtained under conditions of coherent illumination
of a disordered 2D sample. The influence of partial coherence of x-rays on the results of the x-ray cross-correlation analysis
was briefly discussed in Ref.\cite{Altarelli}. Here we will consider this question in more detail.

Partial coherence of x-rays can be characterized in terms of the mutual intensity function\cite{Goodman1, Wolf} $J_{\text{in}}(\mathbf{ r}_{1},\mathbf{ r}_{2})$ of the beam incoming on the sample and defined as $J_{\text{in}}(\mathbf{ r}_{1},\mathbf{ r}_{2})=\langle E(\mathbf{ r}_{1},t)  E(\mathbf{ r}_{2},t)\rangle_{T}$, where averaging is performed over times $T$ much longer than the fluctuation time of the incoming x-ray field. It describes the statistical properties of the wavefield as a correlation function between two values of the electric field, $E(\mathbf{ r}_{1},t)$ and $E(\mathbf{ r}_{2},t)$, at different points $\mathbf{ r}_{1}$ and $\mathbf{ r}_{2}$ in space and at the same time $t$, averaged over fluctuations of the wavefield.
It is convenient also to introduce a normalized complex
coherence factor,
\begin{equation}
\mu_{\text{in}}(\mathbf{ r}_{1},\mathbf{ r}_{2}) = J_{\text{in}}(\mathbf{ r}_{1},\mathbf{ r}_{2})/[I_{\text{in}}(\mathbf{ r}_{1})I_{\text{in}}(\mathbf{ r}_{2})]^{1/2},
\label{PCoh1}
\end{equation}
where $I_{\text{in}}(\mathbf{ r}_{1})$ and $I_{\text{in}}(\mathbf{ r}_{2})$ are the intensity values of the incoming beam at points $\mathbf{ r}_{1}$ and $\mathbf{ r}_{2}$, respectively, averaged over fluctuations of the wavefield.
Using the definition (\ref{PCoh1}), the effect of partial coherence of x-rays on the distribution of the scattered intensity  $I_{\text{pcoh}}(\mathbf{q})$ can be expressed
in the far-field limit
\footnote{Here, we considered a broad distribution of the intensity $I_{\text{in}}(\mathbf{ r}_{1})=I_{\text{in}}(\mathbf{ r}_{2})=\text{const}$ over the size of the sample and a spatially uniform distribution of the complex coherence factor
$\mu_{\text{in}}(\mathbf{ r}_{1},\mathbf{ r}_{2})=\mu_{\text{in}}(\mathbf{ r}_{2}-\mathbf{ r}_{1})$ (Shell-model).}
as a convolution of the coherently scattered intensity $I_{\text{\text{coh}}}(\mathbf{q})$ with the Fourier transform $\mu_{\text{in}}(\mathbf{q})$ of the complex coherence factor\cite{Vartanyants.JSR.2003},
\begin{equation}
I_{\text{pcoh}}(\mathbf{q})=I_{\text{coh}}(\mathbf{q})\otimes\mu_{\text{in}}(\mathbf{q}).
\label{PCoh2}
\end{equation}
One can directly use this equation for calculations of the distribution of the scattered intensity for partially coherent illumination of a sample, as soon as coherence properties of the incoming beam are defined by the known complex coherence factor.

\begin{figure*}[!htbp]
\centering
\includegraphics[width=1.0\textwidth]{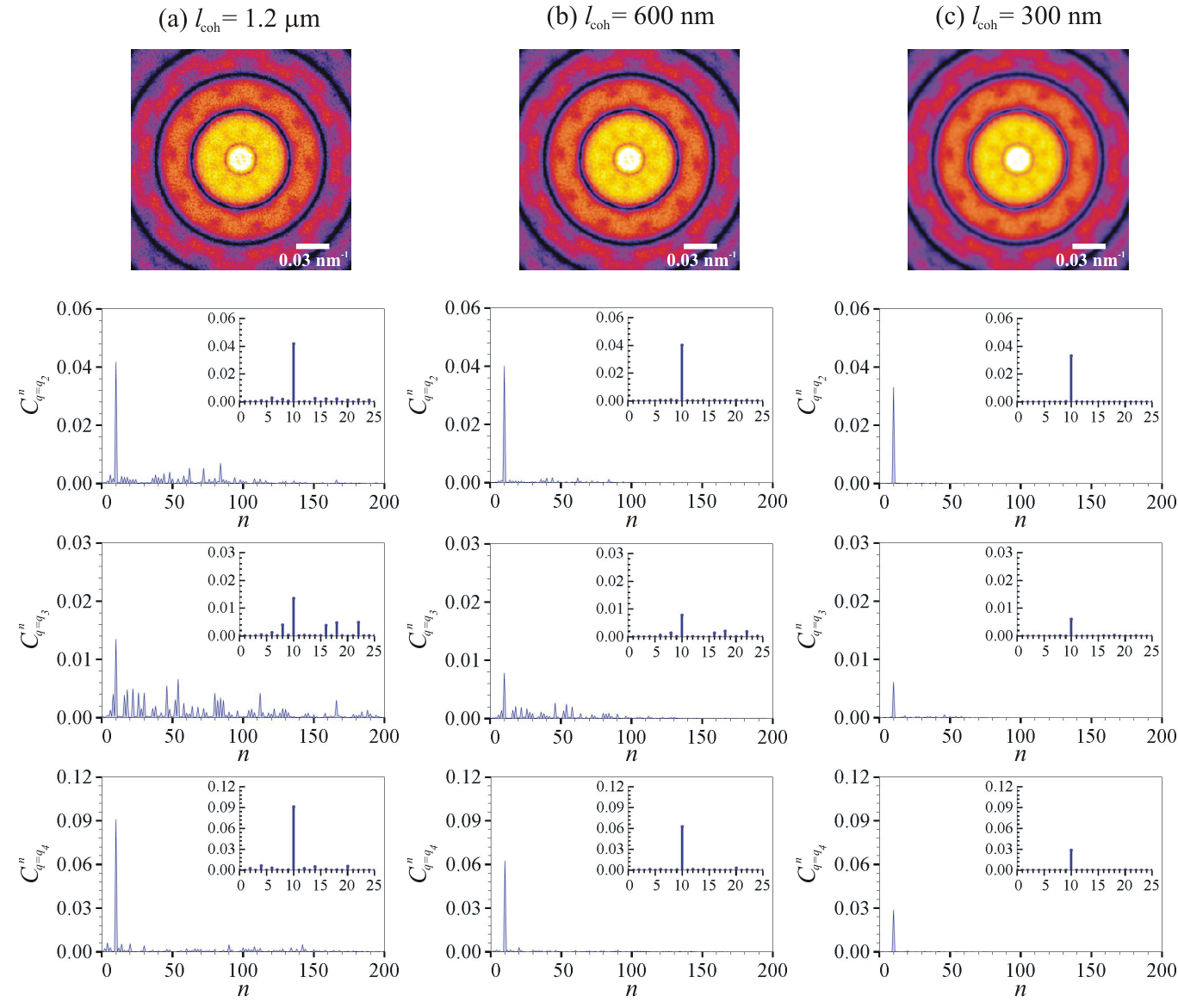}
\caption{\label{Fig:PartCoher}(Color online)
Diffraction patterns and Fourier components of the CCFs calculated for a disordered system consisting of 11 clusters in the same positions and orientations as in the sample 3 in Fig.~\ref{Fig:EnsembleCompare}. Different values of the transverse coherence length $l_{\text{coh}}$ of the incoming x-ray beam: (a) $l_{\text{coh}}=1.2\;\mu\text{m}$, (b) $l_{\text{coh}}=600\;\text{nm}$, and (c) $l_{\text{coh}}=300\;\text{nm}$ were considered. Fourier spectra were calculated for each case at three $q$ values, $q_{2}= 0.029\;\text{nm}^{-1}$,
$q_{3}= 0.036\;\text{nm}^{-1}$, and $q_{4}= 0.043\;\text{nm}^{-1}$.}
\end{figure*}

We will demonstrate the influence of partial coherence of x-rays on the results of the XCCA
by analyzing the scattering experiment with different degree of spatial coherence of the incident x-rays for a disordered system consisting of 11 clusters, presented in Fig.~\ref{Fig:EnsembleCompare}(c) (for Sample 3). We assume in our simulations a Gaussian form of the complex coherence factor (so-called Gaussian Shell-model\cite{Wolf})
\begin{equation}
\mu_{\text{in}}(\mathbf{ r}_{1},\mathbf{ r}_{2}) =\exp[-(\mathbf{ r}_{1}-\mathbf{ r}_{2})^2/2l_{\text{coh}}^{2}],
\label{PCoh3}
\end{equation}
where $l_{\text{coh}}$ is the transverse coherence length.
Three different values of
the transverse coherence length,  $l_{\text{coh}}=1.2\;\mu\text{m}, 600\;\text{nm}$, and $300\;\text{nm}$ are considered
in our calculations.
These values of the coherence length $l_{\text{coh}}$ allow to probe typical length scales in a chosen sample, starting from comparably large coherent length, going down to the size of a single pentagonal cluster and below.
In Fig.~\ref{Fig:PartCoher} diffraction patterns calculated according to Eq.~(\ref{PCoh2},\ref{PCoh3}) as a function
of the transverse coherence length $l_{\text{coh}}$, and  corresponding Fourier spectra of the CCFs calculated at three different $q$ values for each pattern are presented. As one can see from Fig.~\ref{Fig:PartCoher}, decrease of the transverse coherence length corresponds to smearing of speckles in a diffraction pattern.
At the same time, smearing of speckles reduces the contribution from inter-cluster correlations in the Fourier spectra of the CCFs. Partial coherence of the incident x-rays acts as a filter for the Fourier components of the CCF with a threshold defined by the value of the transverse coherence length. Decrease of the value of $l_{\text{coh}}$ down to the size of a single pentagonal cluster considerably reduces the high-frequency contribution in the Fourier spectra, however, the contributions from the inter-cluster correlations  do not vanish completely [compare Fig.~\ref{Fig:PartCoher}(b) with the results for the case of fully coherent illumination of the Sample 3 in Fig.~\ref{Fig:EnsembleCompare}(c)]. Further decrease of the value of $l_{\text{coh}}$ below the size of a single cluster leads to a further
refinement of the Fourier spectra, and the contribution from the local structure becomes clearly dominant [Fig.~\ref{Fig:PartCoher}(c)].

Comparison of the Fourier components of the CCFs $C_{q}^{n}$ ($n=10,20$) for the case of fully coherent scattering ($C_{q}^{n\;\text{coh}}$, $l_{\text{coh}}=\infty$) and partially coherent scattering ($C_{q}^{n\;\text{pcoh}}$, $l_{\text{coh}}=600\;\text{nm}$) from a system of 11 clusters with a uniform distribution of orientations is presented in Table~\ref{Table4}. Note, that absolute values of the Fourier components $C_{q}^{n\;\text{pcoh}}$ corresponding to partially coherent data are significantly lower at all $q$ values, as compared to $C_{q}^{n\;\text{coh}}$ values. At the same time, the inter-cluster contribution is suppressed significantly stronger and as a result one gets higher contrast for the LS contribution.

It is interesting to compare these results with the ones presented in
Ref.~\cite{Vartanyants.JSR.2003}, where the effects of a partial coherence were studied on a model sample represented by an ordered array of identical particles in the same orientation. It was demonstrated\cite{Vartanyants.JSR.2003}, that when the transverse coherence length of the incoming beam illuminating such a sample approaches the size of a single particle, the resulting
diffraction pattern looks similar to the one produced by a single particle. Taking this into account, one would expect that for the disordered sample, considered here, a decrease of the transverse coherence length down to the size of a single pentagonal cluster will lead to a diffraction pattern, which resembles a sum of $11$ diffraction patterns obtained for a single cluster in $11$ different orientations.
Such a diffraction pattern with the partial illumination of 
x-rays could be sufficient for the future phase retrieval analysis\cite{Kam, Kam1, Saldin, Saldin1}.

Analysis of the effect of partial coherence on the results of the cross-correlation analysis allows the following conclusions to be made for disordered 2D systems. If the value of the transverse coherence length is of the order of a size of a single particle (cluster, molecule), the contribution of the inter-particle correlations strongly decreases (completely vanishing in the case of a dilute system), enabling easier identification of the Fourier components related to the internal structure of particles. Therefore, one may prefer to perform a scattering experiment with partially coherent x-ray beam, in order to extract the information on local structure of a disordered system. On the other hand, varying a degree of coherence of x-rays, one could probe the structural correlations in a disordered system at different length scales, which may be particularly useful in studies of a medium range order.


\section{Conclusions}

In a summary, our simulations have demonstrated that in order to get reliable information about the local structures the systems with a high degree of orientational order are preferable ones comparing to the completely disordered systems. Due to the statistical nature of the CCFs the information on local structure of a completely disordered system from a single diffraction pattern is practically not accessible. Averaging of CCFs over an ensemble of diffraction patterns allows to overcome this difficulty and gives reliable information on the structure of clusters forming the system. 
An estimate of a number of diffraction patterns necessary to determine an averaged CCF with a certain accuracy was obtained. Our analysis shows that few tens of thousands of diffraction patterns have to be averaged to get a reliable result.

The limiting factor for this analysis becomes the number of particles in the system and the inter-particle correlations that are always present in the conditions of coherent illumination. Our simulations show that the systems with a small number of clusters are preferable. Averaging of CCFs over an ensemble of diffraction patterns for such systems can significantly enhance the contrast of the structural contribution as compared to the inter-particle contribution. For the systems with a large number of clusters this averaging procedure could become not efficient due to the strongly suppressed signal (it scales as $1/N$ with the number of particles $N$) from the structural contribution.
Our analysis has shown that the averaged inter-particle contribution practically does not depend on the number of particles and their density, but rather depends on the size and the shape of the system as well as on the size of clusters composing the system.
Utilizing partially coherent incident beams with the coherence length about the size of the clusters composing the system could suppress the inter-particle contribution and make possible determination of the local structure of a disordered system.

Interesting open question left for the future work is an extension of XCCA to the three-dimensional (3D) systems. The additional degree of freedom for orientation of 3D clusters introduces a significant complication in the extraction of the structural information from the XCCA analysis. As it was pointed out in our previous work\cite{Altarelli} one possible solution can be the measurements performed at high scattering angles when the curvature of the Ewald sphere has to be taken into account. This will be especially interesting case for atomic systems (organic and non-organic molecules, biological systems, etc.) as opposed to the systems composed of colloidal particles considered in this work. The full cross-correlation analysis for such practically important systems could open new horizons in exploring the nature of disordered systems.


\begin{acknowledgments}

We acknowledge a careful reading of the manuscript by H. Franz and A. Bell.
Part of this work was supported by BMBF Proposal 05K10CHG "Coherent Diffraction Imaging and Scattering of Ultrashort
Coherent Pulses with Matter" in the framework of the German-Russian collaboration "Development and Use of Accelerator-Based Photon Sources".

\end{acknowledgments}

\bibliography{XCCA_simulations_paper_literature}


\appendix

\section{\label{AppA}} 


Here we present the results of calculations of the function $L^{n}(q)$
[Eq.~(\ref{LocStr1})] for a centered
pentagonal 2D cluster [Fig.~\ref{Fig:ExpGeometr}(b)].
The direction of the incident x-ray beam is assumed to be parallel to 5-fold rotational axis of a cluster.
We define the electron density of a cluster as a real-valued quantity in the following form
\begin{equation}
\widetilde{\rho}(\mathbf{r})=\sum\limits _{i=1}^{N_{s}}f_{i}(q)\delta(\mathbf{ r}-\mathbf{ r}_{i}),
\label{AII.Ro}
\end{equation}
where $f_{i}(q)$ and $\mathbf{r}_{i}$ are the form factor and the radius vector of the $i$-th scatterer and $N_{s}$ is the number of scatterers in the cluster. Substituting Eq.~(\ref{AII.Ro}) into Eq.~(\ref{LocStr1}) and assuming that the cluster is composed of scatterers with the same form factor $f(q)$ we obtain
\begin{equation}
L^{n}(q)=\vert f(q)\vert^2\sum_{s,t}J_{n}(q|\mathbf{ r}_{st}|)e^{-in\phi_{\mathbf{ r}_{st}}}\label{AII.Ln}
\end{equation}
Using the coordinates of scatterers in Eq.~(\ref{AII.Ln}) we obtain for a centered pentagonal 2D cluster
\begin{equation}
L^{n}(q)=
\left\{
\begin{array}{l l}
\vert f(q)\vert^2\{6J_{n}(0)+10[J_{n}\left(q a \right)+J_{n}\left(A_{1}q a \right)+J_{n}\left(A_{2}q a \right)]\}& \text{if }n=0,\\
& \quad n \text{ mod } 20=0;\\

10\vert f(q)\vert^2[J_{n}\left(qa\right)-J_{n}\left(A_{1}qa\right)-J_{n}\left(A_{2}qa\right)]& \text{if }n \text{ mod } 10=0,\\
& \quad n \text{ mod } 20\neq0;\\
0 & \text{other } n,\\
\end{array} \right.
\label{AII.Pentagon}
\end{equation}
where $a$ is a distance between centers of the nearest particles inside the pentagonal cluster, $A_{1}=\sqrt{\frac{1}{2}(5-\sqrt{5})}\approx 1.18$, and $A_{2}=\sqrt{\frac{1}{2}(5+\sqrt{5})}\approx 1.9$.
Note, that the function $L^{n}(q)$ presented here for a \textit{centered} pentagonal cluster differs from the one defined in Ref.~\cite{Altarelli} for a LS with a fivefold symmetry due to the missing central particle in the latter case.

In our simulations we consider the following form factor of a spherical particle
\begin{equation}
f(q)=\rho\frac{4\pi[\sin(qr)-(qr)\cos(qr)]}{q^3},
\label{AII.ScatFact}
\end{equation}
where $\rho$ is the average electron density inside a particle and $r$ is the radius of a particle.

\section{\label{AppC}} 


Here we determine a statistical estimate of the average Fourier components of the inter-cluster contribution
$\langle C^{n}_{q\;(int-clust)}\rangle$ [see Eq.~(\ref{AverCorr2b})].
First, we will consider a simplified model of a disordered system, which consist of spherical particles. Next, we will show by simulations that asymptotic equations derived for this simplified model can be used to determine the inter-particle contribution in a disordered system composed of more complicated objects.

According to Eq.~(\ref{Eq:IntClustAver}) the normalized Fourier component $C^{n}_{q\;(int-clust)}$ can be written as follows
\begin{equation}
C^{n}_{q\;(int-clust)} =S^{n}_{4}/[I^{0}(q)]^{2},
\label{C1}
\end{equation}
where
\begin{subequations}
\begin{eqnarray}
&&S^{n}_{4}=4\left\vert \sum\limits_{k_{2}>k_{1}} L_{k_{1},k_{2}}^{n}(q) \right\vert^2,\label{C1a}\\
&&\left\vert I^{0}(q)\right\vert^{2}=\left\vert\sum\limits_{k_{1},k_{2}} L_{k_{1},k_{2}}^{0}(q)\right\vert^2
=\left\vert NL^{0}(q)+2\sum\limits_{k_{2}>k_{1}} L_{k_{1},k_{2}}^{0}(q)  \right\vert^2.\label{C1b}
\end{eqnarray}
\end{subequations}
Here $I^{0}(q)$ was decomposed into contribution of two terms with $k_{1}=k_{2}$ and $k_{1}\neq k_{2}$ similar to the expression in Eq.~(\ref{Inq}), $L_{k_{1},k_{2}}^{n}(q)$ is defined by Eq.~(\ref{Ln_k1k2}), and $L^{0}(q)$ by  Eq.~(\ref{LocStr1}) with $n=0$.

Using the expression for the electron density $\widetilde{\rho}(\mathbf{r})$ [see Eq.~(\ref{AII.Ro})]
in Eqs.~(\ref{Ln_k1k2}) and (\ref{LocStr1}) for a system of clusters composed of identical scatterers we can write~\cite{Altarelli}
\begin{subequations}
\begin{eqnarray}
&&L_{k_{1},k_{2}}^{n}(q)=\sum_{s,t}J_{n}(q|\mathbf{ R}^{s,t}_{k_{2},k_{1}}|)e^{-in\phi_{\mathbf{ R}^{s,t}_{k_{2},k_{1}}}},\label{C2aa}\\
&&L^{0}(q)=\sum_{s,t}J_{0}(q|\mathbf{ r}_{st} |).\label{C2bb}
\end{eqnarray}
\end{subequations}
In Eq.~(\ref{C2aa}) the summation is performed over positions of the scatterers inside the clusters $k_{1}$ and $k_{2}$, and in Eq.~(\ref{C2bb}) inside a single cluster\footnote{Due to normalization the scattering factors cancel out in the expression for $C^{n}_{q\;(int-clust)}$ in Eq.~(\ref{C1}) for identical particles forming clusters.}.

Now, we consider a simplified model of a disordered system, where we substitute a cluster by a spherical particle of the same size as the original cluster.
In this case $\mathbf{ r}_{st}=0$ and Eqs.~(\ref{C2aa}, \ref{C2bb}) reduce to
\begin{subequations}
\begin{eqnarray}
&&L_{k_{1},k_{2}}^{n}(q)=J_{n}(q|\mathbf{ R}_{k_{2},k_{1}}|)e^{-in\phi_{\mathbf{ R}_{k_{2},k_{1}}}},\label{C2a}\\
&&L^{0}(q)=1.\label{C2b}
\end{eqnarray}
\end{subequations}
Further, the double summation in $S^{n}_{4}$ and $I^{0}(q)$ [Eqs.~(\ref{C1a}) and (\ref{C1b})] can be substituted by a single sum over the same set of vectors. Preserving the number of contributions in the original sum
we can write
\begin{subequations}
\begin{eqnarray}
&&S^{n}_{4}=4\lvert \sum\limits_{k=1}^{N_{1}}J_{n}(q|\mathbf{ R}_{k}|)e^{-in\phi_{\mathbf{ R}_{k}}}\rvert^{2},\label{C4a}\\
&&\left[ I^{0}(q)\right]^{2}=\left[ N+2\sum\limits_{k=1}^{N_{1}}J_{0}(q|\mathbf{ R}_{k}|)\right]^{2},\label{C4b}
\end{eqnarray}
\end{subequations}
where $N_{1}=N(N-1)/2$.

The sum in Eq.~(\ref{C4a}) is, in fact, a random phasor sum\cite{Goodman1,Goodman2} defined as
\begin{equation}
\mathbf{B}_{n}=B_{n}e^{i\theta_{n}}=\frac{1}{N_{1}}\sum\limits_{k=1}^{N_{1}}b_{k}^{n}e^{in\phi_{k}},\quad n\neq 0,
\label{C5}
\end{equation}
where $B_{n}$ and $\theta_{n}$ are the length and phase of the random phasor sum $\mathbf{B}_{n}$, and the amplitude and phase of each term are given by
\begin{equation}
b_{k}^{n}\equiv J_{n}(q|\mathbf{ R}_{k}|),\quad  \phi_{k} \equiv \phi_{\mathbf{R}_{k}}.
\label{C6}
\end{equation}
With these definitions we can also formally write the random sum in Eq.~(\ref{C4b}) in the following form,
\begin{equation}
B_{0}=\frac{1}{N_{1}}\sum\limits_{k=1}^{N_{1}}b_{k}^{0}.
\label{C5a}
\end{equation}
Substituting Eqs.~(\ref{C5} - \ref{C5a}) into Eqs.~(\ref{C4a}, \ref{C4b}) we obtain
\begin{subequations}
\begin{eqnarray}
&&S^{n}_{4}=4N_{1}^{2}B_{n}^{2},\label{C7a}\\
&&\left[I^{0}(q)\right]^{2}=\left[N+2N_{1}B_{0}\right]^{2}=N^2+4NN_{1}B_{0}+ 4N_{1}^{2}B_{0}^{2}\label{C7b}.
\end{eqnarray}
\end{subequations}

As soon as we are interested in averaged characteristics of the Fourier component $C^{n}_{q\;(int-clust)}$ [Eq.~(\ref{C1})] we note that both terms $ S^{n}_{4}$ and $[I^{0}(q)]^2$ can be considered as random variables as well. Applying the theorem on a linearized approximation~\cite{Rasch} to evaluate $\langle C^{n}_{q\;(int-clust)} \rangle$ we can express an average of the ratio of these two variables as a ratio of the corresponding averages

\begin{equation}
\langle C^{n}_{q\;(int-clust)} \rangle \approx \langle S^{n}_{4}\rangle/\langle[ I^{0}(q)]^2\rangle,
\label{C3}
\end{equation}
where
\begin{subequations}
\begin{eqnarray}
&&\langle S^{n}_{4}\rangle=4N_{1}^{2}\langle B_{n}^{2}\rangle,\label{C8a}\\
&&\langle[I^{0}(q)]^2\rangle = N^2+4NN_{1}\langle B_{0}\rangle+ 4N_{1}^{2}\langle B_{0}^{2}\rangle.\label{C8b}
\end{eqnarray}
\end{subequations}
It is clear that this approach is valid for $q-$values when $\langle[I^{0}(q)]^2\rangle \neq 0$. Our direct simulations have shown that approximation described by Eq.~(\ref{C3}) is well satisfied for our systems.

According to Eqs.~(\ref{C3}), (\ref{C8a}) and (\ref{C8b}), in order to estimate the value
$\langle C^{n}_{q\;(int-clust)} \rangle$ we need to determine the statistical averages
 $\langle B_{n}^{2}\rangle$ and $\langle B_{0}\rangle$.
Following the approach of Ref.~\cite{Goodman2},
one can obtain the marginal probability density function for the random variable $B_{n}^2$ ($n\neq 0$), that reduces for a large number of
particles to a normal distribution
\begin{equation}
P(B_{n}^2)=\frac{1}{2\sigma_{n}^2}\exp\left( -\frac{B_{n}^2}{2\sigma_{n}^2}\right),\quad B_{n}^2\geq 0,
\label{C12}
\end{equation}
where
\begin{equation}
\sigma^{2}_{n}=\frac{1}{N_{1}^2}\sum_{k=1}^{N_{1}}\frac{\langle(b^{n}_{k})^{2}\rangle}{2}= \frac{\langle J^{2}_{n}(q|\mathbf{R}|)\rangle}{2N_{1}}.
\label{C9}
\end{equation}
Using the probability density function $P(B_{n}^2)$ the mean value of the random variable $B_{n}^2$ and its variance can be determined for $n\neq 0$ as
\begin{subequations}
\begin{eqnarray}
\langle B_{n}^2\rangle&&=2\sigma^{2}_{n}=\langle J^{2}_{n}(q|\mathbf{R}|)\rangle/N_{1},\label{C14a}\\
\sigma^{2}_{B_{n}^2}&&=4\sigma^{4}_{n}=[\langle J^{2}_{n}(q|\mathbf{R}|)\rangle/N_{1}]^2.\label{C14b}
\end{eqnarray}
\end{subequations}
According to Eqs.~(\ref{C14a}) and (\ref{C14b}) the value $B_{n}^2$ fluctuates around the mean value $\langle B_{n}^2\rangle$ with the standard deviation $\sigma_{B_{n}^2}$. According to Eq.~(\ref{C8a}), this also determines the statististical behaviour of the term $S^{n}_{4}$.
Note, that similarly to the random phasor sum considered for LS contribution, averaging over sufficiently large number $M$ of diffraction patterns leads to a decrease of the variance $\sigma^2_{B_{n}^2}$, and the result of such averaging $\langle B_{n}^2\rangle_{M}$ asymptotically approaches its statistical limit, i.e. $\langle B_{n}^2\rangle_{M}\to \langle B_{n}^2\rangle$ (see Appendix~\ref{AppB}).

The average values $\langle B_{0}\rangle$ and $\langle B_{0}^{2}\rangle$ can be directly determined as
\begin{subequations}
\begin{eqnarray}
\langle B_{0}\rangle &&= \frac{1}{N_{1}}\sum_{k=1}^{N_{1}}\langle b^{0}_{k}\rangle=\langle J_{0}(q|\mathbf{R}|) \rangle,\label{C19a}\\
\langle B_{0}^{2}\rangle &&= \frac{1}{N_{1}^2}\sum_{k_{1},k_{2}}\langle b^{0}_{k_{1}} b^{0}_{k_{2}}\rangle=
\frac{1}{N_{1}^2}\left[\sum_{k=1}^{N_{1}}\langle J_{0}^{2}(q|\mathbf{R}_{k}|)\rangle+
\sum_{k_{1}\neq k_{2}}\langle J_{0}(q|\mathbf{R}_{k_{1}}|)J_{0}(q|\mathbf{R}_{k_{2}}|)\rangle\right]\nonumber\\
&&=\frac{1}{N_{1}}\left[\langle J_{0}^{2}(q|\mathbf{R}|)\rangle+
(N_{1}-1)\langle J_{0}(q|\mathbf{R}|)\rangle^{2}\right],\label{C19b}
\end{eqnarray}
\end{subequations}
In the derivation of Eq.~(\ref{C19b}) we assumed that for $k_{1}\neq k_{2}$, $\langle J_{0}(q|\mathbf{R}_{k_{1}}|)J_{0}(q|\mathbf{R}_{k_{2}}|)\rangle=
\langle J_{0}(q|\mathbf{R}_{k_{1}}|)\rangle \langle J_{0}(q|\mathbf{R}_{k_{2}}|)\rangle$.

Using these results in Eqs.~(\ref{C8a}) and (\ref{C8b}) we obtain for the statistical averages $\langle S^{n}_{4}\rangle$ and $\langle[I^{0}(q)]^2\rangle$,
\begin{subequations}
\begin{eqnarray}
&&\langle S^{n}_{4}\rangle=4N_{1}\langle J^{2}_{n}(q|\mathbf{R}|)\rangle,\label{C20a}\\
&&\langle[I^{0}(q)]^2\rangle = N^2+4N_{1}\left[ N\langle J_{0}(q|\mathbf{R}|)\rangle+
\langle J^{2}_{0}(q|\mathbf{R}|)\rangle+
(N_{1}-1)\langle J_{0}(q|\mathbf{R}|)\rangle^2\right].\label{C20b}
\end{eqnarray}
\end{subequations}

For a large number of particles in the system $N \gg 1$, $N_{1}\approx N^{2}/2$ and
substituting now Eqs.~(\ref{C20a}, \ref{C20b}) into Eq.~(\ref{C3}) we obtain for the statistical average
\begin{equation}
\langle C^{n}_{q\;(int-clust)} \rangle =
\frac{2\langle J^{2}_{n}(q|\mathbf{R}|)\rangle}
{1+2N\langle J_{0}(q|\mathbf{R}|)\rangle+2\langle J^{2}_{0}(q|\mathbf{R}|)\rangle+N^{2}\langle J_{0}(q|\mathbf{R}|)\rangle^{2}}.\label{Eq:Cfin1}
\end{equation}
Our direct simulations for the systems with a large number of clusters considered here show, that the terms $N\langle J_{0}(q|\mathbf{R}|)\rangle$, $\langle J^{2}_{0}(q|\mathbf{R}|)\rangle$ as well as $N^{2}\langle J_{0}(q|\mathbf{R}|)\rangle^{2}$
in the denominator of Eq.~(\ref{Eq:Cfin1}) are much smaller than unity and Eq.~(\ref{Eq:Cfin1}) reduces to the following asymptotic expression
\begin{equation}
\langle C^{n}_{q\;(int-clust)} \rangle \approx
2\langle J^{2}_{n}(q|\mathbf{R}|)\rangle.\label{Eq:Cfin2}
\end{equation}
As we can see from this expression, for the systems considered here the inter-cluster contribution 
practically does not depend on the number of clusters $N$.
Substituting Eqs.~(\ref{Eq:Cfin2}) and (\ref{AverCorr3}) in Eq.~(\ref{AverCorr2b}) for a sufficiently large number of diffraction pattern we obtain
\begin{equation}
\langle C^{n}_{q} \rangle =\mathcal{L}^{n}(q)\cdot\langle A^2_n \rangle +
                               2\langle J^{2}_{n}(q|\mathbf{R}|)\rangle.\label{Eq:Cfin3}
\end{equation}
Taking into account that for a uniform distribution of orientations $\langle A^2_n \rangle=1/N$ [Eq.~(\ref{UnifDistr})] we obtain the upper limit for the number of particles $N$, for which the structural contribution
dominates over the inter-cluster contribution. The first term in Eq.~(\ref{Eq:Cfin3}) strongly dominates, if the number of clusters $N$ satisfies the following condition
\begin{equation}
N \ll \frac{\mathcal{L}^{n}(q)}{2\langle J^{2}_{n}(q|\mathbf{R}|)\rangle}.\label{Eq:NumPart}
\end{equation}
For the systems considered here this estimate shows that the number of particles in the system should be considerably lower than about a few hundreds, that is in a good agreement with our observations.

\begin{figure*}[!htbp]
\centering
\includegraphics[width=0.9\textwidth]{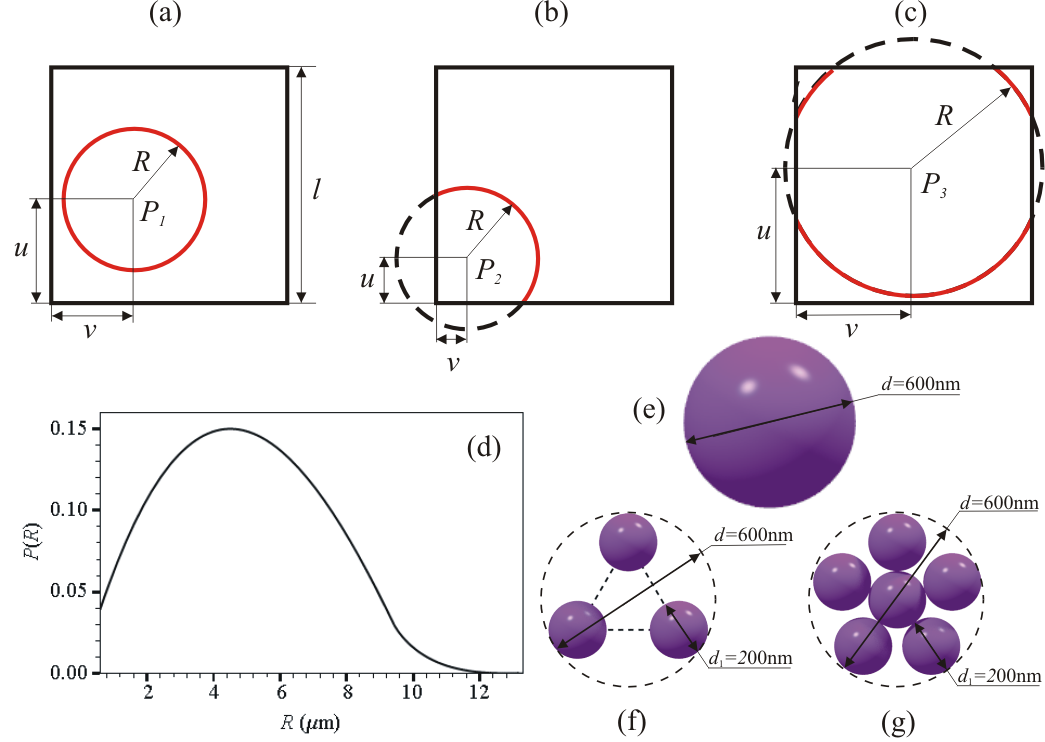}
\caption{\label{Fig:Appendix3_1}(Color online) Derivation of the RDF for a square sample.
(a)-(c) Three cases when the reference particle is placed at the position $P_{i}$ ($i=1,2,3$), which is located from the sides of the square at distances $u$ and $\nu$, are shown. $P_{\text{part}}(R,u,\nu)$ is proportional to (a) the length of the circumference of the radius $R$, (b) the length of the single arc bounded by the square, and (c) the total length of three arcs bounded by the square (in all cases solid red parts of the circumference); (d) RDF for spherical particles of a diameter $d=600\;\text{nm}$ distributed within the square 2D sample with the length of a side of the square $l=10\;\mu\text{m}$; Three model systems used in calculations of inter-particle contribution consist of (e) spherical particles (f) triangular clusters and (g) centered pentagonal clusters.}
\end{figure*}

\bigskip
The averages $\langle J_{0}(q|\mathbf{R}|)\rangle$ and $\langle J^{2}_{n}(q|\mathbf{R}|)\rangle$ (for all $n$ values)
required to calculate $\langle C^{n}_{q\;(int-clust)} \rangle$ can be determined using the radial distribution function (RDF) $P(R)$
\begin{equation}
\langle J_{0}(q|\mathbf{R}|)\rangle=\int_{R_{\text{min}}}^{R_{\text{max}}}P(R)J_{0}(qR)dR,\quad
\langle J^{2}_{n}(q|\mathbf{R}|)\rangle=\int_{R_{\text{min}}}^{R_{\text{max}}}P(R)J^{2}_{n}(qR)dR,\label{C15}
\end{equation}
where $R_{\text{min}}$ and $R_{\text{max}}$ are the minimum and maximum inter-particle distances in the system.
According to its definition\cite{Kodama}, $P(R)$ gives a number of particles in an
annulus of thickness $dR$ at a distance $R$ from another particle. Note, that $P(R)$ defined in such a way should be normalized by the number of particles $N$ before using it in Eq.~(\ref{C15}), in order to satisfy the condition $\int_{R_{\text{min}}}^{R_{\text{max}}}P(R)dR=1$. In our simulations we consider a statistical distribution of spherical particles with a diameter $d$ within the square 2D sample
with the length $l$ of a side of the square. For such a system, it is convenient to determine the RDF using the ``partial radial distribution function''\cite{Kodama} (PRDF) $P_{\text{part}}(R,u,v)$,
\begin{equation}
P(R)=\frac{1}{S_{fs}}\int_{u=d/2}^{l-d/2}\int_{\nu=d/2}^{l-d/2}P_{\text{part}}(R,u,\nu)dud\nu,\quad d\leq R\leq \sqrt{2}(l-d),
\label{CRDF1}
\end{equation}
where $P_{\text{part}}(R,u,\nu)$ defines the number of particles in an
annulus of thickness $dR$ at a distance $R$ from a particle at a
position $P$ which is located from the sides of the square at distances $u$ and $v$
[see Fig.~\ref{Fig:Appendix3_1}], and $S_{fs}=(l-d)^2-\pi d^2$ is a part of the sample area accessible for $P_{\text{part}}(R,u,\nu)$. Depending on the radius $R$ and the position $P$, $P_{\text{part}}(R,u,\nu)$ can be proportional to the length of the circumference [Fig.~\ref{Fig:Appendix3_1}(a)], the length of a single arc [Fig.~\ref{Fig:Appendix3_1}(b)], or the total length of the few arcs [Fig.~\ref{Fig:Appendix3_1}(c)]. Clearly, the RDF depends on the size and the shape of the particles and the sample. Eq.~(\ref{CRDF1}) ensures that the minimum distance between the particles is $d$, and minimum distance from a particle center to the sample edge is $d/2$; it also defines the region of integration in Eq.~(\ref{C15}) with $R_{min}=d$, $R_{max}=\sqrt{2}(l-d)$. The RDF $P(R)$ calculated using Eq.~(\ref{CRDF1}) for a system of spherical
particles with $d=600\;\text{nm}$ and the size of the sample $l=10\;\mu\text{m}$ is presented in Fig.~\ref{Fig:Appendix3_1}(d).
As we can see from this figure, the maximum of this distribution lies between $4\;\mu\text{m}$ and $6\;\mu\text{m}$
that corresponds approximately to the half size of the system.

We compare now the results of an asymptotic estimate $\langle C^{n}_{q\;(int-clust)}\rangle$ of the inter-cluster contribution for a system composed of spherical particles using Eqs.~(\ref{C3}), (\ref{C20a}) and  (\ref{C20b}) with the results of direct calculations for different systems by averaging $C^{n}_{q\;(int-clust)}$ in Eq.~(\ref{C1}) over $M=1000$ realizations. In the latter case, the coordinates of particles in each realization were used in calculations of Eqs.~(\ref{C1a}) and (\ref{C1b}).
Three 2D disordered systems composed of (a) spherical particles [Fig.~\ref{Fig:Appendix3_1}(e)], (b)
triangular clusters [Fig.~\ref{Fig:Appendix3_1}(f)] and (c) centered pentagonal clusters [Fig.~\ref{Fig:Appendix3_1}(g)] were
considered in these direct calculations.

\begin{figure*}[!htbp]
\centering
\includegraphics[width=0.5\textwidth]{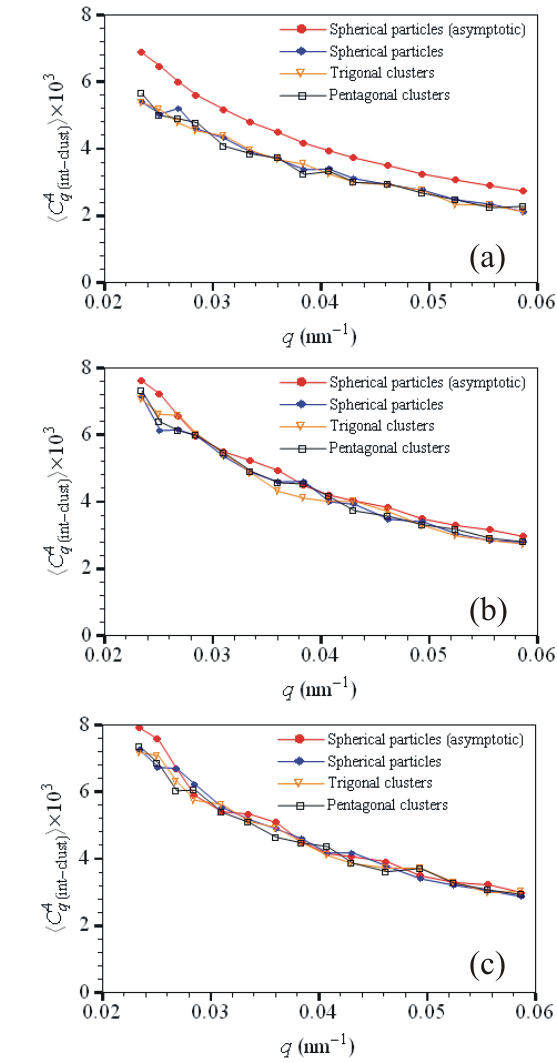}
\caption{\label{Fig:Appendix3_2}(Color online)
Fourier components of the average CCFs corresponding to the inter-particle correlations $\langle C^{n}_{q\;(int-clust)}\rangle$ (for $n=4$).
Calculations were performed using asymptotic expressions [Eqs.~(\ref{C3}), (\ref{C20a}) and  (\ref{C20b})] for a system
composed of spherical particles (red circles). For comparison, the results of direct averaging of Eq.~(\ref{C1}) over the set of $M=1000$  diffraction patterns for a system composed of spherical particles (blue diamonds), trigonal clusters (yellow triangles) and pentagonal clusters (black squares) are shown. The number of particles in the system (a) $N=11$, (b) $N=60$, and (c) $N=121$.}
\end{figure*}

Fourier components $\langle C^{n}_{q\;(int-clust)}\rangle_{M}$ and $\langle C^{n}_{q\;(int-clust)}\rangle$ calculated using Eqs.~(\ref{C1}) and (\ref{C3}) for $n=4$ and different $q$ values for systems with $N=11$, $60$ and $121$ particles are presented in Fig.~\ref{Fig:Appendix3_2}. The Fourier component with $n=4$ was considered because for the chosen symmetry of clusters only the term $S^{n}_{4}$ in Eq.~(\ref{Cq1q2n_2}) contributes to $C_{q}^{n}$ [terms $S^{n}_{1}$ and $(S^{n}_{2}+S^{n}_{3})$ are equal to zero].
In all three cases direct calculations show that the inter-cluster contribution practically does not depend
on the internal structure of particles. As one can see from Fig.~\ref{Fig:Appendix3_2}, the inter-cluster contribution can be accurately estimated using the asymptotic expression.
The inter-cluster contribution calculated using the asymptotic expression [Eqs.~(\ref{C3}), (\ref{C20a}) and  (\ref{C20b})] for three systems considered here for an entire Fourier spectrum is presented in Fig.~\ref{Fig:CorrAver} with the solid red line. As one can see, our estimate quite accurately reproduces the results of direct calculations for all systems considered here in a wide range of $n$ values.
We want to note here, that asymptotic values for the system with $N=11$ clusters slightly overestimate contribution from the inter-cluster correlation [see Fig.~\ref{Fig:CorrAver}(a) and Fig.~\ref{Fig:Appendix3_2}(a)] that is due to the violation of the central limit theorem, used in the derivation of Eq.~(\ref{C3}).

Finally, our analysis shows that the asymptotic expression for the average inter-particle contribution
$\langle C^{n}_{q\;(int-clust)}\rangle$
[Eqs.~(\ref{C3}), (\ref{C20a}) and  (\ref{C20b})] can be effectively used as an estimate of this contribution to
the total Fourier component $\langle C^{n}_{q} \rangle_{M}$ in Eq.~(\ref{AverCorr2b}).
We demonstrated that $\langle C^{n}_{q\;(int-clust)}\rangle$ depends
on the shape and size of the sample, as well as the size of particles, and rather weakly on their number $N$ in the system.


\section{\label{AppB}} 


Here, we give an estimate of the number of diffraction patterns $M$ that is required to determine the ensemble averaged amplitude
$\langle A^{2}_{n} \rangle_{M}$ [Eq.~(\ref{AverCorr4})] with a given accuracy $\varepsilon$.
We define $\varepsilon$ as a maximum deviation of $\langle A^{2}_{n} \rangle_{M}$ from its statistical limit $\langle A_{n}^{2} \rangle$.

Each single x-ray pulse produces a diffraction pattern from a static distribution of clusters in the system, and for the $m$-th diffraction pattern the square modulus $A^{2}_{n}(m)$ of the random phasor sum (\ref{2.24}) can be written as
\begin{equation}
A^{2}_{n}(m)=\langle A^{2}_{n} \rangle + \delta A^{2}_{n}(m),
\label{B1}
\end{equation}
where $\langle A^{2}_{n} \rangle$ is the statistical average and $\delta A^{2}_{n}(m)$ describes fluctuations of this average value from one diffraction pattern to another.

Introducing an average over $M$ diffraction patterns one can write
\begin{equation}
\langle A^{2}(m)\rangle_{M}=\langle A^{2} \rangle + \langle \delta A^{2}(m) \rangle_{M}.
\label{B2}
\end{equation}
Hereafter we omit the subscript $n$ for brevity. The average $\langle A^{2}(m)\rangle_{M}$ can be considered as
a random variable with a variance $\sigma^{2}_{M}$.
We assume, that the measurement of $A^{2}(m)$ is a stationary ergodic process, and $\sigma^{2}_{M\to \infty} \to 0$.
For a sufficiently large but finite number of diffraction patterns $M$ satisfying ergodicity condition, the variance $\sigma^{2}_{M}$ can be estimated as\cite{Papoulis}
\begin{equation}
\sigma^{2}_{M}\simeq\frac{1}{M}\sum\limits_{\Delta m=0}^{M}C(\Delta m).
\label{B5}
\end{equation}
Here $C(\Delta m)$ ($\Delta m=m-m'$) is the autocovariance of $A^{2}(m)$,
\begin{equation}
C(\Delta m)=\langle A^{2}(m) A^{2}(m+\Delta m) \rangle-\langle A^{2} \rangle^{2},
\label{B6}
\end{equation}
where statistical averaging is performed over a large ensemble of diffraction patterns.

Using the estimated value of $\sigma^{2}_{M}$ [Eq.~(\ref{B5})] we can determine a confidence interval for $\langle A^{2} \rangle_{M}$, for a given $\varepsilon$. This can be done applying Tchebycheff's inequality\cite{Papoulis},
\begin{equation}
P\left\{\left\vert \langle A^{2} \rangle_{M}-\langle A^{2} \rangle\right\vert<\varepsilon\right\}>1-\frac{\sigma^{2}_{M}}{\varepsilon^2}.
\label{B9}
\end{equation}
Eq.~(\ref{B9}) determines the probability $P\{x\}$ that the average $\langle A^{2} \rangle_{M}$ lies in the interval $\langle A^{2} \rangle\pm \varepsilon$.

\bigskip

As an example, we consider the system of $N$ clusters with a uniform distribution of orientations $\phi_{k}$ and determine the confidence interval for $\langle A^{2} \rangle_{M}$, for a given accuracy $\varepsilon$. Taking into account that for such a system the probability density function $p[A^{2}(m)]$ is known\cite{Altarelli}
\begin{equation}
p[A^{2}(m)]=Ne^{-NA^{2}(m)},\quad 0\leq A^{2}(m) \leq 1,
\label{B13}
\end{equation}
we obtain for the autocovariance (\ref{B6})
\begin{equation}
C(\Delta m)=
\left\{
\begin{array}{l l}
1/N^2, & \Delta m=0,\\
0, &  \Delta m \neq0.
\end{array} \right.
\label{B15}
\end{equation}
Substituting Eq.~(\ref{B15}) into Eq.~(\ref{B5}) we have for the variance
\begin{equation}
\sigma^{2}_{M}=\frac{1}{M N^2}.
\label{B16}
\end{equation}
Using this value of $\sigma^{2}_{M}$ we can determine the confidence interval for $\langle A^{2} \rangle_{M}$ according to Eq.~(\ref{B9}). For the $10\%$ accuracy $\varepsilon=0.1\langle A^{2} \rangle = 0.1/N$, where we took into account that $\langle A^{2} \rangle=1/N$ for a system considered here. With this we obtain from Tchebycheff inequality [Eq.~(\ref{B9})]

\begin{equation}
P\left\{0.9\langle A^{2} \rangle <\langle A^{2} \rangle_{M} < 1.1\langle A^{2} \rangle\right\}>1-\frac{10^2}{M}.
\label{B17}
\end{equation}
From this inequality the probability $P$ that $\langle A^{2} \rangle_{M}$ lies in the interval from $0.9\langle A^{2}\rangle$ to $1.1\langle A^{2} \rangle$ is $P>0.9$ for $M=10^3$, and $P>0.99$ for $M=10^4$ diffraction patterns.

\newpage

\begin{flushleft}
\begin{table}
\footnotesize
\caption{\label{Table1} Fourier components of the CCFs $C_{q}^{n}$ ($n=10,20$) for a single centered pentagonal cluster [$C_{q}^{n\;\text{dilute}}\equiv \mathcal{L}^n(q)$] and for a completely oriented dense system of clusters ($C_{q}^{n\;\text{dense}}$) described in Section~\ref{Sec4} (Fig.~\ref{Fig:PentagonCase1}). The relative standard deviation is defined as $\delta C_{q}^{n} =(C_{q}^{n\;\text{dense}}-C_{q}^{n\;\text{dilute}})/C_{q}^{n\;\text{dilute}}$.}
\begin {tabular}{|p{5cm}|c|c|c|c|c|c|}
\hline
Model & \multicolumn{5}{|c|}{$C_{q}^{10}$}& $C_{q}^{20}$ \\
\hline
$q,\;[\text{nm}^{-1}]$ & $q_{1}=0.023$ & $q_{2}=0.029$ & $q_{3}=0.036$ & $q_{4}=0.043$ & $q_{5}=0.059$ & $q_{5}=0.059$ \\
\hline
$C_{q}^{n\;\text{dilute}}\equiv \mathcal{L}^n(q)$, [Eq.~(\ref{Lnq})]  & 0.225 & 0.142 & 0.028 & 0.366& 0.143& 0.152 \\
\hline
$C_{q}^{n\;\text{dense}}$, [Fig.~\ref{Fig:PentagonCase1}(d)]  & 0.245 & 0.101 & 0.044 & 0.313& 0.156& 0.175 \\
\hline
$\delta C_{q}^{n}$, \% & 9 & -29& 57& -14& 9& 15\\
\hline
\end{tabular}
\end{table}
\end{flushleft}
\begin{flushleft}
\begin{table}
\footnotesize
\caption{\label{Table2} Fourier components of the CCFs $C_{q}^{n}$ ($n=10,20$) for a dilute
($\langle C_{q}^{n\;\text{dilute}}\rangle$) and dense ($C_{q}^{n\;\text{dense}}$) systems composed of 121 centered pentagonal clusters
with the Gaussian distribution of orientations described in Section~\ref{Sec4} (Fig.~\ref{Fig:PentagonCase2}). The parameter $\delta C_{q}^{n}$ is defined as in Table~\ref{Table1}.}
\begin {tabular}{|p{5cm}|c|c|c|c|c|c|}
\hline
Model & \multicolumn{5}{|c|}{$C_{q}^{10}$}& $C_{q}^{20}$ \\
\hline
$q,\;[\text{nm}^{-1}]$ & $q_{1}=0.023$ & $q_{2}=0.029$ & $q_{3}=0.036$ & $q_{4}=0.043$ & $q_{5}=0.059$ & $q_{5}=0.059$ \\
\hline
$\langle C_{q}^{n\;\text{dilute}}\rangle$, [Eq.~(\ref{2.26})] &
0.152 & 0.096& 0.019 & 0.247 & 0.096 & 0.031 \\
\hline
$C_{q}^{n\;\text{dense}}$, [Fig.~\ref{Fig:PentagonCase2}(d)] &
0.263 & 0.078 & 0.016 & 0.233 & 0.071 & 0.024 \\
\hline
$\delta C_{q}^{n}$, \% &
73 & -18 & -15 & -6 & -26 & -23 \\
\hline
\end{tabular}
\end{table}
\end{flushleft}

\begin{flushleft}
\begin{table}
\footnotesize
\caption{\label{Table3} Fourier components of the averaged CCFs $\langle C^{n}_{q} \rangle_{M}$ ($n=10,20$) for a dilute ($\langle C^{n\;\text{dilute}}_{q} \rangle_{M},\;M\to \infty$) and dense ($\langle C^{n\;\text{dense}}_{q} \rangle_{M}$) systems composed of $N=11$ centered pentagonal clusters with the uniform distribution of orientations. The parameter $\delta C_{q}^{n}$ is defined as in Table~\ref{Table1}.}
\begin {tabular}{|p{5cm}|c|c|c|c|c|c|}
\hline
Model & \multicolumn{5}{|c|}{$\langle C^{10}_{q} \rangle_{M}$}& $\langle C^{20}_{q} \rangle_{M}$ \\
\hline
$q,\;[\text{nm}^{-1}]$ & $q_{1}=0.023$ & $q_{2}=0.029$ & $q_{3}=0.036$ & $q_{4}=0.043$ & $q_{5}=0.059$ & $q_{5}=0.059$ \\
\hline
$\langle C^{n\;\text{dilute}}_{q} \rangle_{M}$, [Eqs.~(\ref{AverCorr3})] &
2.05E-2 & 1.29E-2 & 0.25E-2 & 3.33E-2 & 1.30E-2 & 1.38E-2 \\
\hline
$\langle C^{n\;\text{dense}}_{q} \rangle_{M}$, [Fig.~\ref{Fig:CorrAver}(a)] &
2.56E-2 & 1.67E-2 & 0.56E-2 & 3.56E-2 & 1.37E-2 & 1.49E-2 \\
\hline
$\delta C_{q}^{n}$, \% &
25 & 29 & 120 & 7 & 15& 13 \\
\hline
\end{tabular}
\end{table}
\end{flushleft}
\begin{flushleft}
\begin{table}
\footnotesize
\caption{\label{Table4} Fourier components of the CCFs $C_{q}^{n}$ ($n=10,20$) for the case of a fully coherent scattering [$C_{q}^{n\;\text{coh}}$] and a partially coherent scattering ($C_{q}^{n\;\text{pcoh}}$, $l_{\text{coh}}=600\;\text{nm}$) from a system of 11 clusters with uniform distribution of orientations. The relative standard deviation is defined as $\delta C_{q}^{n} = [C_{q}^{n\;\text{pcoh}}-C_{q}^{n\;\text{coh}}]/C_{q}^{n\;\text{coh}}$.}
\begin {tabular}{|p{5cm}|c|c|c|c|c|c|}
\hline
Model & \multicolumn{5}{|c|}{$C_{q}^{10}$}& $C_{q}^{20}$ \\
\hline
$q,\;[\text{nm}^{-1}]$ & $q_{1}=0.023$ & $q_{2}=0.029$ & $q_{3}=0.036$ & $q_{4}=0.043$ & $q_{5}=0.059$ & $q_{5}=0.059$ \\
\hline
$C_{q}^{n\;\text{coh}}$, [Fig.~\ref{Fig:EnsembleCompare}(c)] & 0.077 & 0.050 & 0.024 & 0.107& 0.045& 0.022 \\
\hline
$C_{q}^{n\;\text{pcoh}}$, [Fig.~\ref{Fig:PartCoher}(b)]  & 0.050 & 0.040 & 0.008 & 0.063& 0.031& 0.019 \\
\hline
$\delta C_{q}^{n}$, \% & -35 & -20& -68& -42& -31& -12\\
\hline
\end{tabular}
\end{table}
\end{flushleft}
\newpage

\end{document}